\documentclass[]{imag-ms-template}

\usepackage[expansion=false]{microtype}
\usepackage{amsmath}
\usepackage{bm}
\usepackage{booktabs}
\usepackage{array}
\usepackage{caption}
\usepackage{subcaption}
\usepackage{graphicx}
\usepackage{float}
\usepackage{placeins}

\usepackage{xcolor}
\newcommand{\stored}[1]{\textcolor{green!50!black}{#1}}
\newcommand{\tiss}[1]{\textcolor{blue!65!black}{#1}}
\newcommand{\scan}[1]{\textcolor{red!70!black}{#1}}

\graphicspath{{figures/}}

\newcommand{\Gt}{\mathbf{G}(t)}

\newcommand{\rw}{\mathbf{r}_w}
\newcommand{\Nw}{N_\mathrm{w}}

\newcommand{\Nq}{N_\mathrm{q}}
\newcommand{\TE}{\mathrm{TE}}
\newcommand{\dmipysim}{\texttt{dmipy-sim}}

\title{Replayable Monte Carlo MRI: From Persistent Tissue Microphysics to Queryable Virtual Phantoms}

\author{Rutger H.J. Fick,$^{1\ast}$\\
{\small $^{1}$dmrai-lab, Independent Research Initiative (\url{https://dmrai-lab.org})}\\
{\small $^\ast$Correspondence: rutger.fick@dmrai-lab.org}
}
\date{}

\begin{document}

\maketitle


\begin{abstract}

Multiphysics Monte Carlo simulation is the reference method for diffusion MRI in idealized tissue: walkers diffuse
through a numerical substrate with relaxation, permeability, susceptibility and magnetization transfer, and
accumulate the signal of a prescribed acquisition. Yet walker histories are massive, petabytes for a tractography-scale phantom, so
current simulators treat them as transient. Thus, they are regenerated for every new acquisition, field strength or
relaxation rate, although they are identical for all of them. I propose to keep these histories instead as a \emph{replay pack},
compressed into persistent, shareable tissue microphysics, so that any such query is answered cheaply as an inner
product against the pack. By storing each path as a Brownian bridge cut to the band a scanner class can play, transform coding its
coefficients and contracting them into time segments, a replay pack is up to 5\,000 times smaller than the histories
it keeps, while staying within the Monte-Carlo floor for every waveform of that class. It matches analytical
solutions in pores, and MC/DC and MCMRSimulator on realistic axon meshes.
Taking replay to the scale of the field's largest Monte Carlo-generated benchmark, I re-simulate DiSCo, the billion-walker Monte Carlo
tractography phantom, as a \emph{multiphysics} replay phantom. It reproduces the published protocol's signal to one
percent, and any waveform, relaxation or susceptibility up to 100~ms is a query of the same replay pack. I release every replay pack of this paper on the SubstrateCommons Hugging Face Hub
and deploy replayable DiSCo as a public Hugging Face Space (\url{https://huggingface.co/spaces/rfick/disco-replay}),
where anyone can choose an acquisition type (PGSE, OGSE, PGSTE or spherical tensor encoding) and the tissue's parameters, and within two minutes
receive their multiphysics Monte Carlo diffusion-weighted image volumes. With replay, a Monte Carlo simulation becomes an instrument anyone can query, at scale.

\end{abstract}

\section{Introduction}\label{sec:intro}

Diffusion MRI measures the thermal motion of water over micrometres. The motion is independent of the scanner, but the signal depends entirely on its encoding: the method ``is not a single biomarker but an
acquisition-and-model framework'' \citep{LeBihan2026}, in which ``two very different
tissues can produce similar ADC values''. The response has been to diversify the acquisition and to make the
biophysical description of tissue more explicit, and the demand for validation of what those descriptions mean
has grown with them \citep{JelescuBudde2017,Novikov2019,LeBihan2026}. Monte Carlo simulation is one of the few
settings where both the microscopic tissue state and the signal are known
\citep{HallAlexander2009camino,FieremansLee2018cookbook}. Walkers are propagated through a numerical substrate and
their magnetization evolved under a sequence, with few assumptions about the substrate, and a generation of
simulators has made that routine: Camino \citep{HallAlexander2009camino}, MC/DC \citep{RafaelPatino2020mcdc},
Disimpy \citep{Kerkela2020disimpy}, the Realistic Microstructure Simulator \citep{Lee2021rms}, and SpinDoctor by
finite elements \citep{Li2019spindoctor}. The substrates have grown with them: axons packed by growth
\citep{Callaghan2020config} or from spheres \citep{Ginsburger2019medusa}, bundles with dispersion, crossings and
myelin \citep{VillarrealHaro2023} and the glia between them \citep{NguyenDuc2026}, axons traced from electron
microscopy with their caliber variation and undulation \citep{Lee2020beading,Lee2020shape,Andersson2020axon},
permeable membranes \citep{NedjatiGilani2017}, up to phantoms at the scale of a tractography benchmark
\citep{RafaelPatino2021,DISCOChallenge2023}; and a current simulator evolves spins through an arbitrary sequence
with diffusion, relaxation, permeability, susceptibility, surface relaxation and magnetization transfer active in
one substrate at once \citep{Cottaar2026mcmr}. A microscopic simulation is now a numerical experiment on tissue
rather than a generator of one diffusion-weighted signal.

That completeness exposes an asymmetry. Generating the stochastic histories is the expensive
step, and almost nothing asked afterwards changes them. A different gradient waveform does not
move the membranes a walker has already met; a different field orientation does not alter a
trajectory generated in a fixed substrate; a different surface relaxivity does not change where
and when a walker touched a wall. Yet the histories are either discarded once a signal is produced or kept as trajectories too large and simulator-specific to serve as a scientific object. Precomputation is well established elsewhere in microstructure imaging: matrix formalisms evaluate restricted diffusion under arbitrary waveforms from a tractable pore's eigenmodes \citep{Callaghan1997,Grebenkov2007,Ozarslan2009mcf,Drobnjak2011}, and dictionaries make model fitting linear \citep{Daducci2015}. Monte Carlo itself has been precomputed as a signal library \citep{Rensonnet2019fingerprint,NedjatiGilani2017}, which keeps the acquisition with it: a new sequence is a new library. Monte Carlo poses a different question: its value lies in substrates and physics with no compact analytical object, so the object kept must come from the stochastic simulation itself and be the physics, not one of its signals.

This paper asks what the output of a Monte Carlo simulation of tissue should be, and answers
that it should be the solved physics, with the signal as a query of it,
\begin{equation*}
\mathrm{substrate}
\xrightarrow[\;\text{one-time compute}\;]{\;\text{one Monte Carlo simulation}\;}
\boxed{\text{persistent microphysics}}
\xrightarrow[\;\text{a query}\;]{\;\begin{subarray}{l}\text{any acquisition}\\[1pt] \text{any scanner}\\[1pt] \text{any tissue}\end{subarray}\;}
S .
\end{equation*}
This sounds like caching, but a simulation is not easily cached. If keeping the histories were enough, DiSCo would have shared its trajectories with its meshes, and a multiphysics simulator would ship each spin's path and what it met beside its signals \citep{Cottaar2026mcmr}; neither does, because a trajectory dump is neither small (DiSCo's own walk, kept at every step, would be 1.2~PB; Table~\ref{tab:pack}), nor readable by anyone else, nor scalable to a brain. A persistent object others can reuse needs three contributions, two theoretical and one infrastructural, and this paper makes them:
\begin{enumerate}
\item \textbf{Encoding.} Which quantities of the simulation an acquisition can never change, so that they and only they are
  stored, and how to compress them into a dense form without losing accuracy for the questions asked of them
  afterwards, with a certificate that says so.
\item \textbf{Replaying.} How every later question, another sequence, another scanner, a rotated substrate, a different tissue,
  is posed in that same compressed space, so that the dense object is queried without ever being expanded
  back into the trajectories it came from.
\item \textbf{Sharing.} How such objects are certified, shared, queried by anyone and extended in place, so a substrate's persistent microphysics is computed once and never again.
\end{enumerate}

These contributions rest on a contract. A stored microphysics is reusable only inside a precise vocabulary of
what was simulated: which pools of which substrate a walker can occupy, which walls it can meet and with what
properties, and which sources produce the field it samples, so that the simulation is set up in exactly one way
from its description; how that record is represented, so that every later question knows what it holds and what
may be done with it; and the same again one level up, from one stored substrate to an image built of many. I write
these contracts down as an open specification independent of any simulator, one document each for the substrate,
the stored microphysics, the acquisition and the image \citep{FickRpkSpec2026}; what follows is one implementation
of it. This paper is therefore not another simulator. A replay pack sits after the simulation: any engine that saves its walkers' paths can write one, and the engine of Sec.~\ref{sec:methods:engine} is the one whose channels could be checked one by one against closed forms and two other simulators \citep{Cottaar2026mcmr,RafaelPatino2020mcdc}.

\begin{figure}[!htbp]
\centering
\includegraphics[width=\linewidth]{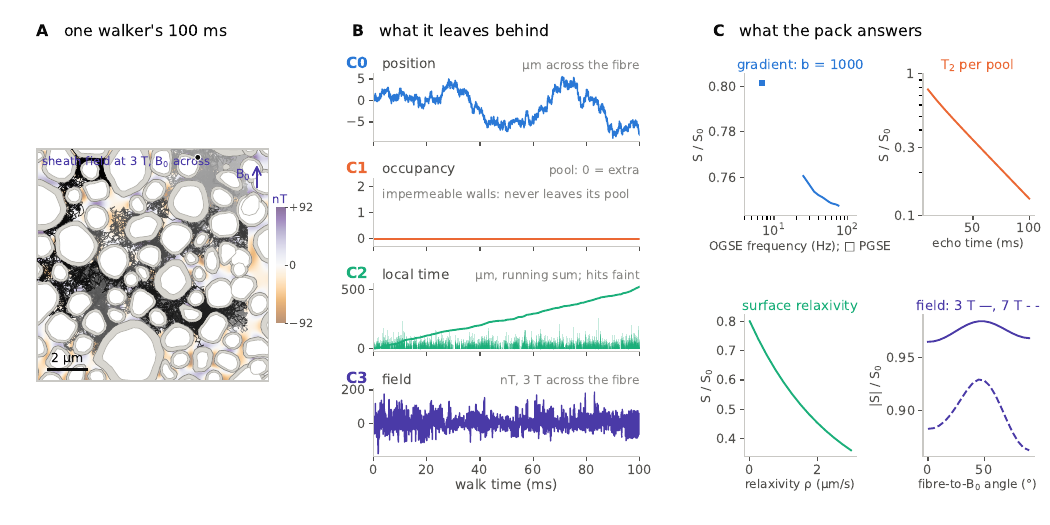}
\caption{One simulation, every acquisition (\texttt{figures/fig\_one\_walk.py}; the CACTUS pack of
Sec.~\ref{sec:results:meshes}). (A) The bundle in cross-section, myelin sheaths grey, and one
extra-cellular walker's 100~ms in it, light to dark with time.
(B) What that walker leaves in the pack: its position across the fibre, its pool, its boundary
local time (the running sum, hits beneath it), and the sheaths' field it passed with
$B_0$ across the fibre, each against simulation time and each the quantity a knob later multiplies.
(C) What the pack answers without another simulation, each curve coloured by the channel it reads: PGSE (at its diffusion-time equivalent $1/4(\Delta-\delta/3)$) and cosine OGSE with one
period per block against frequency at $b=1000$~s/mm$^2$, the spin echo through the pools' $T_2$, PGSE against surface
relaxivity, and the gradient echo against the fibre-to-$B_0$ angle at 3 and 7~T.}
\label{fig:abstract}
\end{figure}

I introduce the \emph{replay pack} as that stored microphysics: for every walker, its path as a Brownian bridge on
a sine basis \citep{Caflisch1997bridge,Glasserman2003mc}, its compartment occupancy, boundary local time and sampled susceptibility field, four series a later acquisition contracts against (Fig.~\ref{fig:abstract}). A pack scales to an image as a \emph{replay phantom}, one simulation of a whole volume with its walkers grouped by voxel, similar to software phantoms that build a voxel from a fibre configuration \citep{Neher2014fiberfox,Caruyer2014phantomas} and benchmark tractography \citep{MaierHein2017challenge}, but with Monte Carlo substrates rather than analytic signal models.
Every pack and phantom reported here, DiSCo's multiphysics reproduction included, is served and queryable under the \href{https://huggingface.co/SubstrateCommons}{\texttt{SubstrateCommons}} organisation on the Hugging Face Hub, each described by a substrate card. Queried, a phantom becomes something new to Monte Carlo simulation: on a public Hugging Face Space, \href{https://huggingface.co/spaces/rfick/disco-replay}{\texttt{disco-replay}}
\citep{DiscoZeroSpace2026}, a visitor chooses an acquisition and, within two minutes on a shared
GPU, receives DiSCo imaged with it, fitted and tracked to a connectome against its ground truth (Sec.~\ref{sec:results:pages}).

The paper builds the claim one layer at a time, each resting on the one before, so the final result -- queryable,
browser-based virtual phantoms -- stands on nothing not shown first.
\begin{enumerate}
\item \textbf{The signal is an inner product with what the replay pack stored} (Sec.~\ref{sec:theory:invariant}).
  Every later choice is a \emph{replay knob}, set after the simulation; what changes the walkers' positions (another geometry, a slower diffusivity) still costs a simulation.
\item \textbf{The scanner sets how much of a walker's path must be stored} (Secs.~\ref{sec:theory:bridge}--\ref{sec:theory:segments},
  \ref{sec:results:band}). The truncation error has a closed form, so a pack certifies which
  scanners it serves; the bits of each kept mode follow from the Monte-Carlo floor by another, and the certified
  pack is 210 times smaller than the trajectory, 5\,000 times once contracted at a timing (Table~\ref{tab:pack}).
\item \textbf{Replay matches closed forms in pores and on a permeable membrane} (Secs.~\ref{sec:results:canonical},
  \ref{sec:results:c1c3}).
\item \textbf{Replay matches two other simulators on realistic axons} (Sec.~\ref{sec:results:meshes}).
\item \textbf{A re-simulated DiSCo reproduces its published images} (Sec.~\ref{sec:results:phantoms}), and answers
  four acquisitions it was never scanned with. Making it a multiphysics phantom took substrate work of its own:
  strands as curved myelinated cylinders on the published centrelines, one wall test per segment in place of some
  24 triangles, and the sheaths' field summed over the whole millimetre by a closed-form segment kernel on a
  particle--mesh grid shipped with the substrate.
\item \textbf{Anyone can query the phantom in a web browser} (Sec.~\ref{sec:results:pages}).
\end{enumerate}
The division of labour is Le Bihan's \citep{LeBihan2026}: the diffusion is the tissue's, and the scanner an
experiment designed to observe it. A replay pack stores the tissue's side once, and every acquisition becomes a
question asked of it. The open question then lies on the substrate side: how far generated micro- and mesoscopic
substrates, replayed this way, predict what a scan shows.

\section{Theory}\label{sec:theory}

\begin{figure}[!tbp]
\centering
\includegraphics[width=\linewidth]{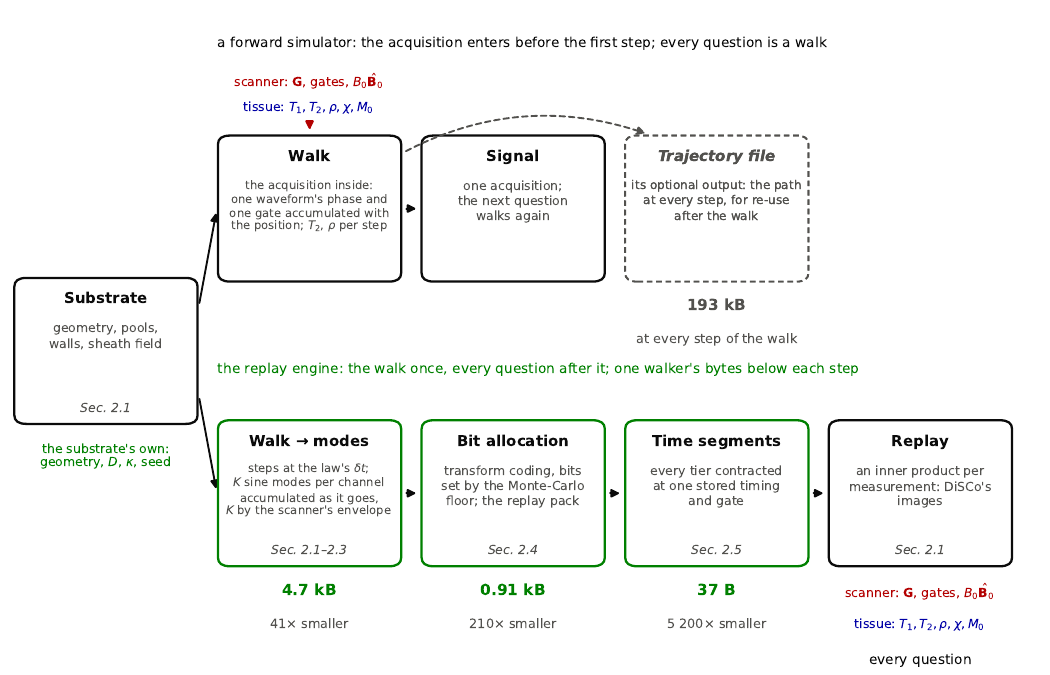}
\caption{One substrate, two roads (\texttt{figures/fig\_pipeline.py}). Above, a forward simulator: the acquisition
enters before the first step and a signal for it comes out, so every question is a walk; its trajectory file, the path at
every step, is the optional output a few of them offer. Below, the replay engine: the walk accumulates the band's modes
as it runs, the transform coding and the time-segment layout follow once, and every question is an inner product at the
end, with the scanner's and the tissue's numbers supplied then. Beneath each step, what one walker costs after it over
100~ms, every tier, and how much smaller that is than the trajectory file at every step (Table~\ref{tab:pack}).}
\label{fig:pipeline}
\end{figure}

The theory follows the road from a substrate to the DiSCo page. The replay
equation writes the signal of one walker as four inner products, one per channel, between series the simulation
stored and functions the tissue and the scanner supply, and draws the boundary of what a stored simulation can
answer (Sec.~\ref{sec:theory:invariant}). What a replay pack answers, over which envelope and to which error, is
stated once (Sec.~\ref{sec:theory:envelope}). Four steps decide what a replay pack keeps of each series, each adding its own
term to that error, drawn in Fig.~\ref{fig:pipeline} and
gathered per channel in Table~\ref{tab:tiers}: the bridge (Sec.~\ref{sec:theory:bridge}), the band
(Sec.~\ref{sec:theory:band}), the bits (Sec.~\ref{sec:theory:bits}) and the time-segment layout
(Sec.~\ref{sec:theory:segments}). Each step is worked in full on
the path, C0, and then for what differs in C1 to C3. Symbols are coloured by owner throughout: the substrate in \stored{green}, the tissue in \tiss{blue},
the scanner in \scan{red}.

\begin{table}[!tbp]
\centering\footnotesize\setlength{\tabcolsep}{4pt}\renewcommand{\arraystretch}{1.2}
\caption{The four channels through the four steps: the envelope that bounds what each answers
(Sec.~\ref{sec:theory:envelope}), what each series is stored as, what sets how many of its
coefficients a pack keeps, the weight $S$ that sets their bits in Eq.~\eqref{eq:container} with the bits that
follow on the CACTUS pack at the 1\% floor, and what one timing contracts it to.}
\label{tab:tiers}
\begin{tabular}{@{}*{6}{>{\raggedright\arraybackslash}p{0.15\linewidth}}@{}}
\toprule
Channel & Envelope (Sec.~\ref{sec:theory:envelope}) & Bridge (Sec.~\ref{sec:theory:bridge}) & Band (Sec.~\ref{sec:theory:band}) & Bits (Sec.~\ref{sec:theory:bits}) & At one timing (Sec.~\ref{sec:theory:segments}) \\
\midrule
C0, path \stored{$\rw$} & the scanner class: $G_{\max}$, $S_{\max}$ per axis & endpoints and $K$ sine modes of the bridge & the waveform: $K=2fT$, a closed-form tail & $S=\gamma^2\sum_kA_k^2$; 11 bits falling to 4 & $\mathbf m_w$: 3 numbers \\
C1, occupancy \stored{$o_{w,i}$} & none: any $T_1$, $T_2$ & events, exact in time & none & none, exact & $\Theta^\perp_{w,j}$, $\Theta^\parallel_{w,j}$: 2 per pool \\
C2, contact \stored{$L_w$} & the largest relaxivity $\rho_{\max}$ & endpoints and $K_L$ sine modes of the bridge & the band of C0 & $S=\sum_k(\omega_k\hat c_k/D)^2$, flat in $k$; 11 bits falling to 3 & $\mathcal L^\perp_w$, $\mathcal L^\parallel_w$: 2 numbers \\
C3, field \stored{$h_w$, $\mathsf M^I_w$, $\mathsf M^A_w$} & the largest $B_0\chi$; $n_\mathrm p\le K_\chi/2$ pulses & thirteen series on $K_\chi$ cosine modes & the gate: $K_\chi\ge2n_\mathrm p$, then a ladder on a pilot & $S\le(\gamma B_0)^2T\sum_cw_c^2$; 9 bits on the first mode, 3 to 4 after & $\Psi_w$: 13 numbers \\
\bottomrule
\end{tabular}
\end{table}

\FloatBarrier
\subsection{The replay equation}\label{sec:theory:invariant}

A Monte Carlo tissue simulation advances walkers by the position update
$d\mathbf r=\sqrt{2D(\mathbf r)}\,d\mathbf W$ with reflection or transmission at membranes
\citep{HallAlexander2009camino,RafaelPatino2020mcdc,Cottaar2026mcmr}, and advances each
walker's magnetization along the path it produces. The two updates are not symmetric. The path
depends on the geometry, the diffusivity $D$, the membrane permeability and the seed, and on
nothing else: no gradient, field, pulse, relaxation time or relaxivity enters it. Everything the
magnetization update needs, it reads off the path. A stored path with its contact record therefore suffices for every magnetization evolution on that substrate, and a
quantity is a \emph{replay knob} if it enters only the magnetization update
(Table~\ref{tab:knobs}).

For any train of instantaneous pulses, the magnetization update is a sum over coherence pathways \citep{Hennig1991echoes,Weigel2015epg}: a pathway $p$ is the sequence of states, transverse, conjugated or stored, the magnetization takes between pulses. Four numbers describe it: its amplitude $\eta_p$, a product of the pulses' transition coefficients that depends on the flip angles alone; its two gates over time, $\chi^\perp_p$ while transverse and $\chi^\parallel_p$ while stored; and its coherence sign $\varepsilon_p=s_p\chi^\perp_p$, $\pm1$ while transverse, flipped by each refocusing pulse, and $0$ while stored. Along one pathway a walker's transverse magnetization at the readout is $M_0\,e^{\Lambda^p_w}$, with $\Lambda^p_w$ a sum of four inner products over time, $\langle f,g\rangle=\int_0^{T}f(t)\,g(t)\,dt$, one per channel:
\begin{equation}
\begin{aligned}
  \Lambda^p_w =\;& \phantom{-}\,i\tiss{\gamma}\,\Big\langle \scan{\varepsilon_p\,\mathbf g(\mathbf x_v,t)},\;
     \stored{\rw}\Big\rangle && \text{C0, position} \\[3pt]
  & -\,\textstyle\sum_i\Big\langle \dfrac{\scan{\chi^{\perp}_p}}{\tiss{T_{2,i}}}+\dfrac{\scan{\chi^{\parallel}_p}}{\tiss{T_{1,i}}},\;
     \stored{o_{w,i}}\Big\rangle && \text{C1, occupancy} \\[3pt]
  & -\,\dfrac{1}{\tiss{D}}\,\Big\langle \tiss{\rho_2}\,\scan{\chi^{\perp}_p}+\tiss{\rho_1}\,\scan{\chi^{\parallel}_p},\;\stored{\ell_w}\Big\rangle && \text{C2, contact} \\[3pt]
  & +\, i\tiss{\gamma}\,\Big\langle \scan{\varepsilon_p},\;
     \scan{B_0}\big[\tiss{\chi_I}\big(\stored{h_w}-\scan{\hat{\mathbf B}_0^{\!\top}}\stored{\mathsf M^I_w}\scan{\hat{\mathbf B}_0}\big)
     +\tiss{\chi_A}\,\scan{\hat{\mathbf B}_0^{\!\top}}\stored{\mathsf M^A_w}\scan{\hat{\mathbf B}_0}\big]
     + \scan{\Delta B(\mathbf x_v)}\Big\rangle && \text{C3, field}
\end{aligned}
\label{eq:channels}
\end{equation}
such that the signal is the walker average over the pathways,
\begin{equation}
  S \;=\; \tiss{M_0}\,\Big|\Big\langle \textstyle\sum_p
  \scan{\eta_p}\,e^{\Lambda^p_w}\Big\rangle_w\Big|,
  \qquad \langle f\rangle_w=\frac{\sum_w \stored{\varpi_w} f_w}{\sum_w \stored{\varpi_w}},
\label{eq:signal}
\end{equation}
\begin{table}[!tbp]
\centering\footnotesize\setlength{\tabcolsep}{4pt}\renewcommand{\arraystretch}{1.15}
\caption{The symbols of Eqs.~\eqref{eq:channels} and \eqref{eq:signal} by owner: what each is, its domain, shape
and unit, and the channel it enters. The \stored{substrate's} series are the pack, written once per simulation,
all functions of where the walker was; the \tiss{tissue's} numbers and the \scan{scanner's} functions are given
at replay.}
\label{tab:symbols}
\begin{tabular}{@{}p{0.13\linewidth}p{0.40\linewidth}p{0.33\linewidth}p{0.08\linewidth}@{}}
\toprule
Symbol & What it is & Domain, shape, unit & Enters \\
\midrule
\stored{$\rw(t)$} & the walker's path & $\mathbb R^3$ in m over $[0,T]$ & C0 \\
\stored{$o_{w,i}(t)$} & occupancy of compartment $i$, the bound pool one of them & $\{0,1\}$ over time, kept as the instants it changes & C1 \\
\stored{$\ell_w(t)$} & the rate at which the walls push the walker back, its local time & $\mathbb R_{\ge0}$, m\,s$^{-1}$ & C2 \\
\stored{$h_w$, $\mathsf M^I_w$, $\mathsf M^A_w$} & the sheath's field along the path per unit of $B_0\chi$, in tensor form: its local term and one tensor per kind of susceptibility & $\mathbb R$, $\mathrm{Sym}(3)$, $\mathrm{Sym}(3)$: thirteen scalar series & C3 \\
\stored{$\varpi_w$} & the walker's weight, its pool's water fraction & $\mathbb R_{\ge0}$ & $\langle\cdot\rangle_w$ \\
\midrule
\tiss{$\gamma$} & gyromagnetic ratio of the nucleus observed & rad\,s$^{-1}$\,T$^{-1}$ & C0, C3 \\
\tiss{$T_{2,i}$, $T_{1,i}$} & relaxation times of compartment $i$ & $\mathbb R_{>0}$, s & C1 \\
\tiss{$\rho_2$, $\rho_1$} & surface relaxivities of transverse and of stored magnetization, acting per unit contact as $\rho/D$ & $\mathbb R_{\ge0}$, m\,s$^{-1}$ & C2 \\
\tiss{$D$} & the pool's simulated diffusivity & $\mathbb R_{>0}$, m$^2$\,s$^{-1}$ & C2 \\
\tiss{$\chi_I$, $\chi_A$} & the sheath's iso- and anisotropic susceptibility & $\mathbb R$, dimensionless & C3 \\
\tiss{$M_0$} & proton density & $\mathbb R_{\ge0}$ & $S$ \\
\midrule
\scan{$\mathbf g(\mathbf x_v,t)$} & the gradient the voxel receives: the prescribed $\mathbf G(t)$ on the ideal scanner, with what a machine adds at first order across the voxel & $\mathbb R^3$ over time, T\,m$^{-1}$ & C0 \\
\scan{$\chi^\perp_p$, $\chi^\parallel_p$} & pathway $p$'s transverse gate and its gate stored along $\hat{\mathbf B}_0$, whatever the bore's axis & $\{0,1\}$ over time & C0--C3 \\
\scan{$q_\nu$, $\hat{\mathbf g}_\nu$, $\psi(t)$} & measurement $\nu$ played as one temporal shape along one direction, $\varepsilon_p\mathbf g=q\,\hat{\mathbf g}\,\psi(t)$: its amplitude, direction and shape, refocusing signs included & T\,m$^{-1}$; the unit sphere; dimensionless & C0, Eq.~\eqref{eq:shape} \\
\scan{$s_p$} & the conjugation sign, flipped by each refocusing pulse; $\varepsilon_p=s_p\chi^\perp_p$ is the coherence sign and $\varepsilon_p\mathbf g$ the effective gradient & $\{-1,+1\}$ over time & C0, C3 \\
\scan{$B_0$, $\hat{\mathbf B}_0$} & the field and its direction, the latter through the quadratic forms of the two tensors & $\mathbb R_{>0}$, T; the unit sphere & C3 \\
\scan{$\Delta B(\mathbf x_v)$} & every field uniform across the voxel, zero on the ideal scanner & $\mathbb R$, T; one phase per voxel, no walker series & C3 \\
\scan{$\boldsymbol\alpha$, $\eta_p$} & the flip angles, which set the gates and the pathway amplitudes $\eta_p$ & rad; $\mathbb C$ & $S$ \\
\bottomrule
\end{tabular}
\end{table}
Magnetization transfer is in these rows and needs no row of its own. MCMRSimulator holds it as stuck particles,
walkers bound to a membrane for a dwell time and relaxing there at their own rates \citep{Cottaar2026mcmr}; here
such a walker occupies the bound pool, one more column of \stored{$o_{w,i}$} with its own \tiss{$T_{2,i}$} and
\tiss{$T_{1,i}$}, so C1 carries it while it stays bound, the binding and release being the walk's
(Table~\ref{tab:knobs}). Its per-collision form, signal lost at every collision, is C2's surface relaxivity. The scanner enters through the gradient $\mathbf g(\mathbf x_v,t)$ and the field offset
$\Delta B(\mathbf x_v)$ the voxel receives; a real scanner's terms in them vary over millimetres while a walker moves
microns, so each multiplies the stored series and never changes them. In this work the scanner is ideal,
$\mathbf g=\mathbf G$, the prescribed waveform, and $\Delta B=0$, and pulses are instantaneous; a real scanner's
terms and finite pulses replay too, and are the Discussion's (Sec.~\ref{sec:discussion:image}). C0 and C3 are
phases: they scale with the field and depend on the substrate's pose, since $\mathbf g$ and $\hat{\mathbf B}_0$ are
in the scanner's frame and the series in the substrate's, so a rotated substrate is a rotated acquisition. C1 and C2 are
magnitudes: they only shorten the magnetization, know nothing of orientation, and are gated only by whether
the magnetization is transverse or stored. 

A spin echo is one pathway, $\varepsilon=(+1,-1)$ with $\eta=1$, so Eq.~\eqref{eq:channels} is its closed
form with $T=\TE$; a stimulated echo is the pathway $\varepsilon=(+1,0,-1)$ with $\chi^\parallel=1$ over the mixing
time, where the occupancy term relaxes at $T_1$ and the position and field terms accrue nothing; a train of $n$ pulses sums over its pathways, whose gates are the same few sign patterns however long it runs.

\begin{table}[!tbp]
\centering\small
\caption{The replay boundary. What enters only the magnetization update is a replay knob, set at replay by the
\scan{scanner} or the \tiss{tissue}; what enters the position update is the \stored{substrate's} and fixed in the
simulation, except through two exact rescalings of diffusion (Sec.~\ref{sec:discussion:limits}).}
\label{tab:knobs}
\begin{tabular}{@{}p{0.34\linewidth}p{0.13\linewidth}p{0.5\linewidth}@{}}
\toprule
Quantity & Set per & Replayable \\
\midrule
\multicolumn{3}{@{}l}{\emph{Replay knobs}} \\
\scan{Gradient waveform, $b$, direction, $\TE\le T_\mathrm{max}$} & acquisition & yes, any deliverable waveform \\
\scan{Pulse train: gates, flip angles, finite pulses} & acquisition & yes; finite pulses by the vector Bloch update \\
\scan{Field strength and direction $B_0\hat{\mathbf B}_0$} & acquisition & yes \\
\scan{Substrate pose} & acquisition & yes, exactly: rotate $\Gt$ and $\hat{\mathbf B}_0$ instead \\
\scan{The scanner's own terms} & scanner, voxel & yes, through $\Delta B$ and $\mathbf g$ at the voxel \\
\tiss{$T_1$, $T_2$} & compartment & yes \\
\tiss{Surface relaxivities $\rho_2$, $\rho_1$} & wall & yes \\
\tiss{Bound-pool $T_1$, $T_2$, offset} & wall & yes \\
\tiss{Susceptibilities $\chi_I$, $\chi_A$} & tissue & yes \\
\tiss{Nucleus: $\gamma$} & tissue & yes: scales the phase rows; its $D$ by a rescaling \\
\tiss{Proton density $M_0$} & tissue & yes: a multiplier \\
\midrule
\multicolumn{3}{@{}l}{\emph{Fixed in the simulation}} \\
\stored{Geometry; box wrap rule} & --- & no: a new simulation \\
\stored{MT binding kinetics} & wall & no: a new simulation \\
\tiss{Diffusivity $D$} & compartment & faster only, as a time scaling \\
\tiss{Permeability $\kappa$} & wall & with $D$, at no other pair \\
\stored{Substrate size $L$} & substrate & as a length scaling, with $D$ and $\kappa$ \\
\bottomrule
\end{tabular}
\end{table}
\FloatBarrier
\subsection{The envelope: what a replay pack answers, and to which error}\label{sec:theory:envelope}

A replay pack does not keep the walk; it keeps the series of Eq.~\eqref{eq:channels} compressed, and its
certificate is a statement about that compression: every signal replayed from it lies within a stated error of the
signal its own walk would give, for every question inside a stated envelope.

The envelope has two sides (Table~\ref{tab:tiers}). On the scanner's side it is a class, the largest gradient
amplitude $G_{\max}$ and slew rate $S_{\max}$ per axis, and the envelope covers every waveform the class can play,
at any $b$-value and direction. On the tissue's side it is the largest value of each knob that scales a stored
series: the surface relaxivity, which multiplies the contact, and $B_0\chi$, which multiplies the field, at any
direction of $B_0$. The relaxation times and the proton density need no bound; they are weights at most one on
series stored exactly. A question outside the envelope is not certified.

Inside it the error is of two kinds, which behave differently:
\begin{itemize}
\item \textbf{Statistical}, the floor $f$. Every knob multiplies a walker's contribution by a phase or by a weight
  at most one, so the standard error of a replayed signal, in units of the voxel's unrelaxed water signal, is at most
  $\sqrt{\sum_w\varpi_w^2}/\sum_w\varpi_w$ whatever the knobs, and the bare signal at high $b$ reaches it. $f$ is a
  standard error, set by the walkers, held for every combination of knobs at once, and lowered by more walkers or by
  averaging over a region. The rounding of the stored coefficients (Sec.~\ref{sec:theory:bits}) belongs here: to
  first order it adds to each walker's phase an error of zero mean, independent of every other walker's, so it widens
  the floor in quadrature as fewer walkers would.
\item \textbf{Bias}, $\varepsilon_\mathrm{num}$, lowered by neither: the truncation of each series above the band its
  reader sees (Sec.~\ref{sec:theory:band}), the walk's quadrature of each coefficient (Sec.~\ref{sec:theory:bridge}),
  its sampling of the field along the path, and the rounding's second order, far below the floor. Each has a closed
  form or is read on a pilot, and their sum is held below $f$.
\end{itemize}
A replay pack states $f$, with the rounding's share in it, beside each term of $\varepsilon_\mathrm{num}$; the steps
that follow fix what a replay pack keeps and derive the term each adds.

\FloatBarrier
\subsection{The bridge: each series as coefficients}\label{sec:theory:bridge}

The \emph{replay pack} stores each series of Eq.~\eqref{eq:channels} as coefficients on a fixed basis in time, on
which every inner product of Eq.~\eqref{eq:channels} is a contraction. The walk accumulates those coefficients on the
fly, so the coefficients, not a trajectory, are the simulation's output.

\paragraph{C0, the path as a Brownian bridge.} A walker's path is stored as its two endpoints
plus the residual between them, a Brownian bridge \citep{Caflisch1997bridge,Glasserman2003mc}, expanded on the sines $s_k(t)=\sqrt{2/T}\sin(\omega_kt)$, $\omega_k=\pi k/T$, orthonormal on $[0,T]$. The waveform's net area and its first moment read the endpoints; the rest of the waveform reads the sines, in order of frequency, so keeping the first $K$ modes keeps the path up to a frequency, and Sec.~\ref{sec:theory:band} sets how many a scanner needs. Gathered on the basis
$\mathbf b(t)=\big(1,\,t,\,s_1(t),\dots,s_K(t)\big)$, a walker is one matrix, a waveform one matrix of the same shape, and the phase their inner product:
\begin{alignat}{3}
  \stored{\rw(t)}&=\stored{\mathsf C_w}\,\mathbf b(t),
  &\quad\stored{\mathsf C_w}&=\Big[\stored{\rw(0)},\;\stored{\big(\rw(T)-\rw(0)\big)}/T,\;\stored{\boldsymbol\beta_{w,1}},\dots,\stored{\boldsymbol\beta_{w,K}}\Big],
  &&\;\text{the stored path}
  \label{eq:bridge}\\
  \varphi_w&=\tiss{\gamma}\,\big\langle\stored{\mathsf C_w},\scan{\mathsf A}\big\rangle,
  &\scan{\mathsf A}&=\int\scan{\varepsilon_p\mathbf g(t)}\,\mathbf b(t)^{\top}dt,
  &&\;\text{any encoding}
  \label{eq:ch:phase}
\end{alignat}
$\mathsf C_w$ and $\mathsf A$ are $3\times(K{+}2)$, $\langle\cdot,\cdot\rangle$ sums the products of their matching
entries. The truncation at $K$ is the approximation that
remains, and the pack's certificate carries it (Sec.~\ref{sec:theory:band}).

\paragraph{C1 to C3, each on the basis its shape asks for.} A walker rarely changes pool, so its occupancy, C1,
is stored as events: the instants it changes pool, exact to the step, with the pool it enters, and a single label
behind impermeable walls. With \scan{$C^\perp(t)$} and \scan{$C^\parallel(t)$} the time before $t$ with the
transverse, respectively stored, gate on, the relaxation is exact on the runs $(i,t_a,t_b)$ between events,
\begin{equation}
  \Lambda^{\mathrm{C1}}_w=-\sum_{\stored{\text{runs }(i,t_a,t_b)}}
  \Big[\frac{\scan{C^\perp(t_b)-C^\perp(t_a)}}{\tiss{T_{2,i}}}+\frac{\scan{C^\parallel(t_b)-C^\parallel(t_a)}}{\tiss{T_{1,i}}}\Big].
  \label{eq:ch:c1}
\end{equation}

The contact, C2, measures how much of the walk the wall stopped: each reflection adds twice the depth
the step would have gone past the wall along its normal, so a head-on hit counts fully and a grazing one hardly at
all. Summed, it is the distance the wall has pushed the walker back, the local time of the reflected path
\citep{Grebenkov2007}, a length, and $\rho/D$ times it is the log of the magnetization the wall leaves
\citep{BrownsteinTarr1979}. A membrane's crossing has the same shape, with probability $\kappa/D$ times the
same increment \citep{Powles1992}. The
cumulative local time $L_w(t)=\int_0^t\ell_w$ is monotone and near linear, so it is stored like Eq.~\eqref{eq:bridge}: its two endpoints and $K_L$ sine modes
\stored{$\tilde\beta_{w,k}$} of its residual. The C2 row integrates the gate
$c(t)=\tiss{\rho_2}\scan{\chi^\perp(t)}+\tiss{\rho_1}\scan{\chi^\parallel(t)}$ against the increment of $L_w$, and
the increment of a sine mode is a cosine, $\dot s_k=\omega_k\sqrt{2/T}\cos(\omega_kt)$; with
$\scan{\hat c_k}=\sqrt{2/T}\int_0^Tc\cos(\omega_kt)\,dt$ the gate's cosine coefficients,
\begin{equation}
  \Lambda^{\mathrm{C2}}_w=-\frac{1}{\tiss{D}}\Big[\frac{\stored{L_w(T)-L_w(0)}}{T}\int_0^T\!c\,dt
  +\sum_{k=1}^{K_L}\omega_k\,\scan{\hat c_k}\,\stored{\tilde\beta_{w,k}}\Big],
  \label{eq:ch:c2}
\end{equation}
two scalars and one $K_L$-term product per walker; ungated, it is $-(\rho_2/D)\big(L_w(T)-L_w(0)\big)$, read off
the endpoints.

The field, C3, is stored the way a diffusion tensor stores diffusivity. At a walker's position the sheath
shifts the field along $\hat{\mathbf B}_0$ by $\hat{\mathbf B}_0^{\!\top}\mathsf M\hat{\mathbf B}_0$ per unit of
$B_0\chi$, as $\hat{\mathbf g}^{\top}\mathsf D\hat{\mathbf g}$ gives the diffusivity along any gradient
\citep{Basser1994}: one symmetric tensor per kind of susceptibility, $\mathsf M^I_w$ for the isotropic and
$\mathsf M^A_w$ for the anisotropic, holds the field for every direction of $B_0$ at once, and where the walker sits
sets it; in a straight sheath's lumen $\mathsf M^A$ lies across the fibre, the $\sin^2\theta$ of
\citet{WhartonBowtell2012}. With the sheath's local term $h_w$, a third of its indicator and of the trace of
$\mathsf M^I_w$, these are thirteen series, functions of the geometry alone, read along the path at an
interval the field's largest gradient sets and stored as their leading $K_\chi$ cosine coefficients, orthonormal on
$[0,T]$, \stored{$\hat h_{w,j}$, $\hat{\mathsf M}^{I}_{w,j}$,
$\hat{\mathsf M}^{A}_{w,j}$}. The field only ever enters through a gate, so the gate is expanded on the same
basis, \scan{$\hat\varepsilon_j$}, and each series' gated integral is one product,
$\Psi^{h}_w=\tiss{\gamma}\sum_{j<K_\chi}\scan{\hat\varepsilon_j}\,\stored{\hat h_{w,j}}$ for the
local term and likewise for the six components of each tensor, $\boldsymbol\Psi^I_w$ and $\boldsymbol\Psi^A_w$.
The field's strength, its two contrasts and its direction are applied last,
the direction as a quadratic form,
\begin{equation}
  \varphi^{\chi}_w=\scan{B_0}\Big[\tiss{\chi_I}\big(\Psi^{h}_w-\scan{\hat{\mathbf B}_0^{\!\top}}\boldsymbol\Psi^{I}_w\scan{\hat{\mathbf B}_0}\big)
  +\tiss{\chi_A}\,\scan{\hat{\mathbf B}_0^{\!\top}}\boldsymbol\Psi^{A}_w\scan{\hat{\mathbf B}_0}\Big],\qquad\text{one gate}
  \label{eq:ch:susc}
\end{equation}
These are the \emph{gated field integrals}, thirteen per walker and gate, the field's phase in tensor form, the same value to rounding as the
$K_\chi$-term sum. C2 and C3 are thus read alike, each through the gate's cosine coefficients: the contact because
the increments of its sines are cosines, the field because it is stored on them.

\FloatBarrier
\subsection{The band: how many modes the scanner sees}\label{sec:theory:band}

\begin{figure}[!tbp]
\centering
\includegraphics[width=\linewidth]{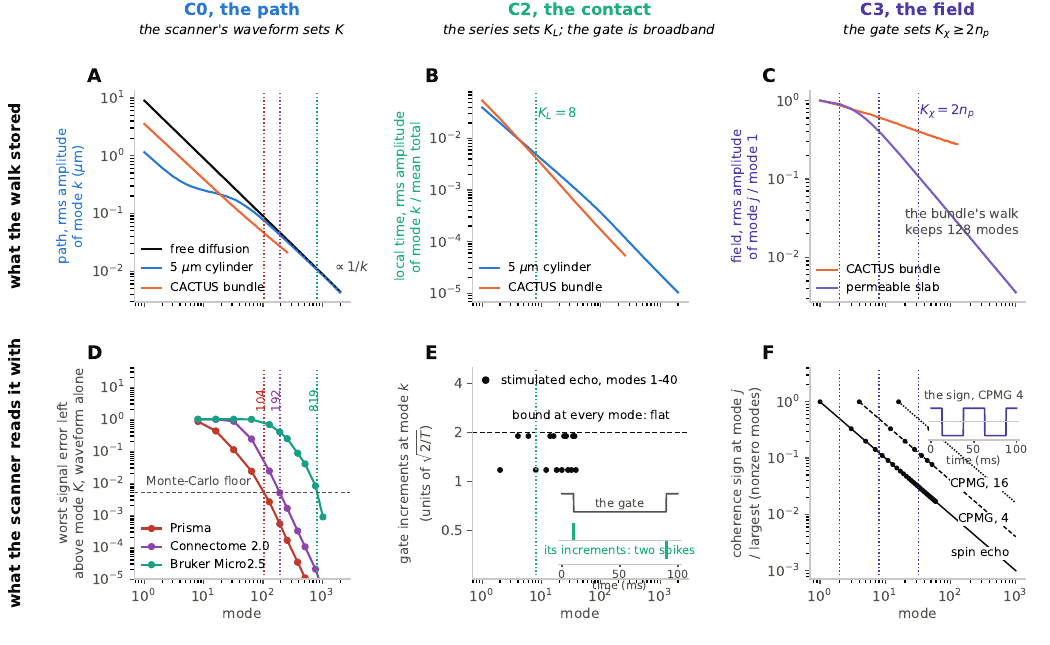}
\caption{Each stored series above the function of the scanner's that reads it, on the same modes
(\texttt{figures/fig\_spectra.py}, data \texttt{fig\_spectra\_data.py}).
\textbf{Top}, the rms amplitude of each mode over the walkers. (A) The path across the fibres: free diffusion and a
5~$\mu$m cylinder walked for 100~ms, 20\,000 walkers each, every mode to 2\,000, and the CACTUS bundle from its
pack, its 256 stored modes. (B) The bridge residual of the cumulative local time over its mean total, the cylinder
and the bundle. (C) The field, over its mode 1: the bundle's isotropic series, its 128 stored modes, and the
permeable slab of Sec.~\ref{sec:results:c1c3}, whose field is the indicator of its off-resonant layer.
\textbf{Bottom}. (D) Per scanner class, the worst signal error a deliverable waveform leaves above mode $K$, from
the waveform alone (Eq.~\eqref{eq:tail}, as in Fig.~\ref{fig:band}), against the floor $\varepsilon=5\times10^{-3}$;
the dotted lines are its crossings, carried up into A. (E) The sine coefficients of a stimulated echo's gate
increments, 80~ms mixing, modes 1 to 40, and their bound at every mode; inset, the gate and its two switches.
(F) The cosine coefficients of the coherence sign of a spin echo and of CPMG trains of 4 and 16 pulses, the nonzero
modes; inset, the 4-pulse sign; the dotted lines are $K_\chi=2n_\mathrm p$, carried up into C.}
\label{fig:spectra}
\end{figure}

Each series above is cut at a number of modes, and each row of Eq.~\eqref{eq:channels} reads it through a
function of the scanner's: the waveform for C0, the gate's switches for C2, the gate for C3. Fig.~\ref{fig:spectra}
sets each series above the function that reads it, on the same modes. The band a channel
needs is the narrower of the two spectra, the series' or the scanner's; when the scanner's is the narrow one the
band follows from the scanner class by a closed form, otherwise it is measured. Measured means one
thing for every row. Let $S_a$ be the ensemble mean of that row's factor alone, the other rows at their identity,
under acquisition $a$, and $S_a^{K}$ the same from the series cut at $K$ modes; over a battery $\mathcal A$ of the
class's shapes, each at the largest amplitude the class allows,
\begin{equation}
  \epsilon_K=\max_{a\in\mathcal A}\big|S_a^{K}-S_a\big|
  \label{eq:certificate}
\end{equation}
is the truncation's term of $\varepsilon_\mathrm{num}$ (Sec.~\ref{sec:theory:envelope}), each row certified on its own
factor. The walk keeps no series to measure $\epsilon_K$ against, so a
pack states $\varepsilon_\mathrm{num}$ as a sum of terms that need none: the truncation, from Eq.~\eqref{eq:tail}
below for C0, from the bands beyond the $K_L$ it stores at the largest relaxivity for C2, and from a pilot for C3; and
the walk's quadrature and its read of the field, each in closed form. A subset of walkers kept whole measures
$\epsilon_K$ directly where a check is wanted.

\paragraph{C0, cut by the waveform, with a closed form.} Mode $s_k$ of Eq.~\eqref{eq:bridge} completes $k$
half-periods in $T$, so $K$ modes resolve the path up to
\begin{equation}
  f_K=\frac{K}{2T}.
  \label{eq:band}
\end{equation}
The path carries every frequency up to its step rate. The waveform does not: a ramp at the slew limit $S_{\max}$
takes $G_{\max}/S_{\max}$ to reach the amplitude limit $G_{\max}$, and an oscillation at $f$ has a quarter
period per ramp, so above $S_{\max}/4G_{\max}$ the amplitude a waveform can reach falls as $S_{\max}/4f$ (Fig.~\ref{fig:spectra}A, D). Thus a
pack keeps $K=2fT$ for the band $f$ of the fastest scanner it serves, and a
consumer reads only the leading $K$ its own hardware needs. The cost of the cut has a closed form to leading order. Split $\varphi_w$ of
Eq.~\eqref{eq:ch:phase} into $\varphi_{w,K}$ and $\varphi_{w,>K}$ for the kept and the discarded modes. For free diffusion the residual of
Eq.~\eqref{eq:bridge} is a discrete Brownian bridge whose Karhunen--Lo\`eve modes are exactly the $s_k$
\citep{Caflisch1997bridge,Glasserman2003mc}, making the $\stored{\boldsymbol\beta_{w,k}}$ independent Gaussian
contributions, so the tail only multiplies the signal; in a pore the residual is not Brownian, and the same holds
approximately above the frequency at which the walls act, where the path moves as if free (the cylinder of
Fig.~\ref{fig:spectra}A). Then
\begin{equation}
  \big\langle e^{i\varphi_w}\big\rangle
  =\big\langle e^{i\varphi_{w,K}}\big\rangle\,e^{-\sigma_{>K}^2/2},
  \qquad
  \sigma_{>K}^2=\frac{2}{\pi}\!\int_{\omega>\pi K/T}\!\!|\mathbf F(\omega)|^2D(\omega)\,d\omega
  \;\le\;2\,D_0\,b_{>K}(\mathbf g),
  \label{eq:tail}
\end{equation}
with $\mathbf F(\omega)$ the Fourier transform of the encoding $\mathbf q(t)=\tiss{\gamma}\int_0^t\scan{\varepsilon_p\mathbf g}\,dt'$,
$D(\omega)$ the diffusion spectrum \citep{Stepisnik1993,CallaghanStepisnik1995}, whose real part never exceeds
$D_0$, the largest diffusivity a walker can reach, and $b_{>K}$ the $b$-value of the waveform above the band,
evaluated on the discarded $s_k$ themselves. The cut loses at most a factor $e^{-D_0b_{>K}}$, read off the
waveform alone (Sec.~\ref{sec:results:band}).

\paragraph{C1 to C3, cut by the series or the gate.} C1's events are exact and have no
band. C2 and C3 are read through the gate, and the gate is broadband: its switches are steps, so $\omega_k\hat c_k$
of Eq.~\eqref{eq:ch:c2} does not fall with $k$, and the coherence sign is a square wave whose harmonics fall only
as $1/j$ (Fig.~\ref{fig:spectra}E, F). The scanner's spectrum is then the wide one, and the series or the gate sets
the band. The two differ in their series. The local time is a cumulative count, near linear, so its bridge residual
carries little (Fig.~\ref{fig:spectra}B), and the walk accumulates it at the band of C0, where its small high modes
cost few bits (Sec.~\ref{sec:theory:bits}). The field a walker samples is broadband (Fig.~\ref{fig:abstract}B): it
crosses the structure of a sheath's field within one save, so its coefficients $\stored{\hat h_{w,j}}$ hardly fall
with $j$ (Fig.~\ref{fig:spectra}C), and no $K_\chi$ small enough to store approximates the series itself. The gate
sets the cut instead. A train of $n_\mathrm p$ pulses switches $n_\mathrm p$ times in $T$, a square wave with
nothing below its cosine mode $n_\mathrm p$, so a pack keeps $K_\chi\ge2n_\mathrm p$, the first mode at half the
band, and declares $n_\mathrm p=K_\chi/2$ as the depth it serves. What the cut loses above $K_\chi$ falls as
slowly as the gate's harmonics and has no closed form, since nothing in it is Gaussian, so it is measured before the
walk: on a pilot of a thousand walkers, a doubling ladder from 16 modes to the band of C0, stopped at the first rung
whose error, Eq.~\eqref{eq:certificate}, is within its share of $\varepsilon_\mathrm{num}$, fixes $K_\chi$ for the walk. The phase the cut drops is $\scan{B_0}\tiss{\chi}$ times a number the walk
fixes, so the pilot runs at the envelope's largest $B_0\chi$, which covers every weaker field and susceptibility.

\FloatBarrier
\subsection{The bits: rounding as a share of the Monte-Carlo floor}\label{sec:theory:bits}

\begin{figure}[!tbp]
\centering
\includegraphics[width=\linewidth]{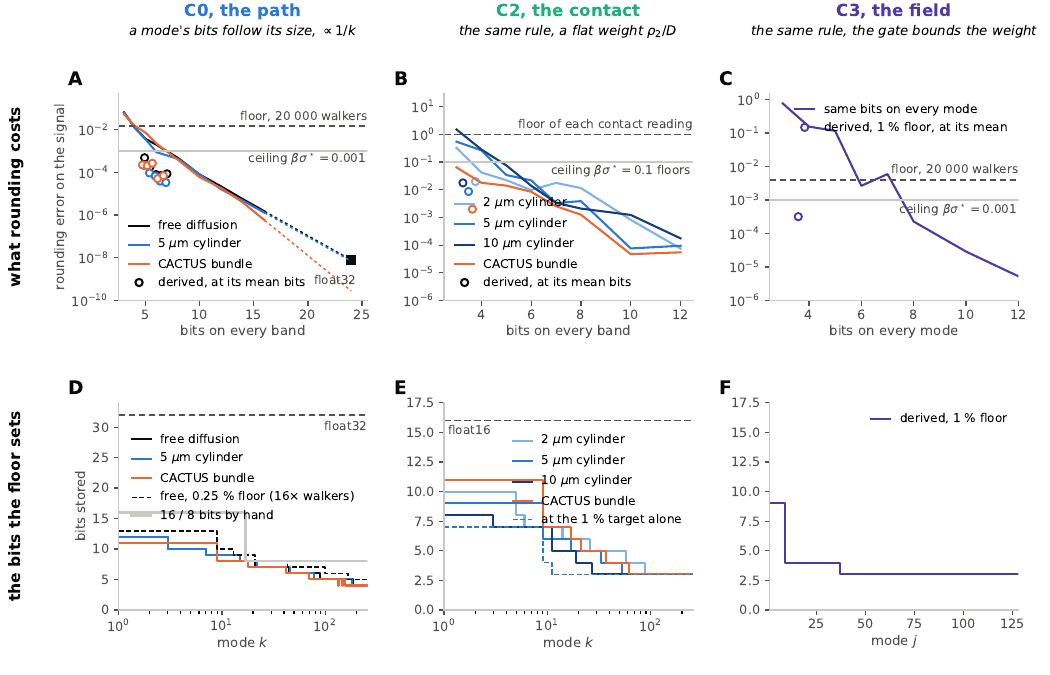}
\caption{The bits of each series, set by the floor (\texttt{figures/fig\_bits.py}, data \texttt{fig\_bits\_data.py}
and \texttt{fig\_container\_data.py}); one column per series, as Fig.~\ref{fig:spectra}. Top: the rounding's own
error on the replayed signal, for the same bits on every coefficient (lines), beside the walkers' Monte-Carlo floor
(dashed) and the ceiling $\beta\sigma^\star$ at the 1\% floor (grey). (A) The path of free diffusion, the
5~$\mu$m cylinder and the CACTUS bundle, 20\,000 walkers, $K=256$, against the float pack on the certificate's
battery, with the container of Eq.~\eqref{eq:container} derived for five floor targets from 2\% to 0.125\%
(circles, at their mean bits) and float32's 24-bit significand (square, extrapolated). (B) The contact series of
2, 5 and 10~$\mu$m cylinders and the CACTUS bundle, the same bits on every band: the worst surface factor over
windows of the walk and relaxivities up to $10^{-4}$~m\,s$^{-1}$, in floors of each contact reading; the circles
are the derived containers. (C) The field of the CACTUS bundle, 20\,000 walkers, its twelve stored series at
$K_\chi=128$, the same bits on every mode: the worst gated signal over a gradient echo, a spin echo, an eight-pulse
train and a stimulated echo, at 3 and 7~T, $B_0$ along and across the bundle; the circle is the derived container. Bottom: the bits Eq.~\eqref{eq:container} gives
each mode at the 1\% floor, beside the precision stored without it. (D) The path, and for free diffusion at 0.25\%,
sixteen times the walkers (dashed); the grey rule is a container chosen by hand, 16 bits on the first sixteen bands
and 8 on the rest. (E) The contact series. (F) The field.}
\label{fig:bits}
\end{figure}

Sec.~\ref{sec:theory:band} fixed how many coefficients each series keeps; it left open how precisely each must be
stored. The coefficients are real numbers, and float32 stores every one of them to a relative $6\times10^{-8}$, far below what a replayed signal can resolve. Rounding
a stored coefficient perturbs what its series contributes to each walker: the phase of Eq.~\eqref{eq:ch:phase} for
the path, the contact term of Eq.~\eqref{eq:ch:c2}, the field's phase of Eq.~\eqref{eq:ch:susc}. The perturbation
has zero mean and is independent from walker to walker, so to first order it is noise of the same kind as the walk's
own: it averages over the walkers and widens the floor $f$ of Sec.~\ref{sec:theory:envelope} in quadrature, as fewer
walkers would. Only its second order is a bias, a term of $\varepsilon_\mathrm{num}$. The coefficients may therefore
be rounded until the floor they widen is a small fraction above the walkers' own, at the lowest floor the replay
pack will be read at (Fig.~\ref{fig:bits}A--C). Matching the
bits spent on each coefficient to its size is transform coding
\citep{HuangSchultheiss1963,GershoGray1992,GrayNeuhoff1998}: one step for every coefficient and $\log_2$ of its size
over that step in bits \citep{Bennett1948}, as JPEG sets a step per frequency \citep{Wallace1991jpeg} and functional
quantisation rounds the Karhunen--Lo\`eve coordinates of Brownian motion \citep{LuschgyPages2002}; a coefficient
that is small over every walker gets few bits (Fig.~\ref{fig:bits}D--F). What is new is the distortion: not a
mean-square error on the series, but the widening of the replayed signal's floor.

\paragraph{C0, the rule.} A coefficient $k$ kept as an integer at its own scale $a_k$, its largest value over the
walkers, with a step $\delta$ common to the coefficients of a series, carries a rounding error uniform on
$\pm\delta/2$. Every series is read linearly, so a walker's rounding errors add into one random perturbation of
what the series contributes, of variance $\epsilon^2=S\,\delta^2/12$, with $S$ the sum of the squared weights by
which the scanner reads the coefficients, taken at its largest over the envelope. For the path it is the
waveform's weight on the bands, $S=\gamma^2\sum_k A_k^2$ in the units of Eq.~\eqref{eq:ch:phase}, and the scanner
class bounds it for every waveform at once: by Bessel's inequality on the orthonormal sines, a waveform with
$|G_\mathrm{axis}|\le G_{\max}$ over $T$ has $\sum_kA_k^2\le G_{\max}^2T$ on each axis, whatever its shape, $b$-value
or direction. To first order the perturbation adds to the mean signal a random term $\kappa\,\epsilon$ times the floor,
$\kappa$ of order one (Sec.~\ref{sec:methods:pack}), so the floor widens from $f$ to $f\sqrt{1+\kappa^2\epsilon^2}$; to second order
it biases the mean by $\epsilon^2/2$, the mean of a cosine over a zero-mean phase, which does not fall with the walker
count while the floor does. A budget $\beta$ of the floor target $\sigma^\star$ caps both, the widening at
$\kappa\epsilon\le\beta/\sqrt2$ and the bias at $\epsilon^2/2\le\beta\sigma^\star/2$, and gives one step and the
bits of every coefficient in closed form,
\begin{equation}
  \epsilon=\min\!\Big(\frac{\beta}{\sqrt2\,\kappa},\;\sqrt{\beta\,\sigma^\star}\Big),\qquad
  \delta=\sqrt{12}\,\frac{\epsilon}{\sqrt S},\qquad
  b_k=\Big\lceil\log_2\frac{a_k}{\delta}\Big\rceil+1\ \text{bits},
  \label{eq:container}
\end{equation}
clipped to 3 to 16 bits and grouped into runs of equal bit count that fill whole bytes. At $\beta=0.1$ the
rounding widens the floor by at most a quarter of a percent, what half a percent fewer walkers would cost. The first
branch binds above
a floor of $\beta/2\kappa^2$; below it the bias sets the step, $\delta\propto\sqrt{\sigma^\star}$, so a replay pack gains
half a bit per band for every halving of its floor, that is for every fourfold increase in walkers, and never
spends a bit below the noise the walkers bought. The path's coefficients fall as $1/k$, and its bits fall with
them, from 11 to 4 over the band (Fig.~\ref{fig:bits}D). Nothing in the container is chosen: the band comes from the
scanner, the bits from the walkers. The floor target is that of the pack as it will be read. A walk run as $M$ shards of the same
voxels and merged is one ensemble with a floor $\sqrt M$ below a shard's, and the bias does not average away, so
$\sigma^\star$ is the union's floor and the bits are derived once, before the shards walk: a quarter of $\log_2M$
bits per band more than one shard alone would take.

\paragraph{C1 to C3, the same rule with their own weights.} C1's events are exact and leave nothing to round. For
the contact the weight is the gate's, $S=\sum_k(\omega_k\hat c_k/D)^2$ of Eq.~\eqref{eq:ch:c2} at the largest
relaxivity, flat in $k$, since each switch of the gate reaches every mode alike (Fig.~\ref{fig:spectra}E); the local
time's coefficients fall nonetheless, and its bits with them, from 11 to 3 (Fig.~\ref{fig:bits}E), about a ninth of
the bytes of float32. For the field it is $S=(\gamma B_0)^2\sum_j\hat\varepsilon_j^2\sum_c w_c^2$, with $w_c$
the weights by which Eq.~\eqref{eq:ch:susc} combines the thirteen series, at the largest $B_0$ and
susceptibilities and the worst direction; the gate is a sign, so by Parseval $\sum_j\hat\varepsilon_j^2\le
T$, whatever its pulses: the broadband gate that sets $K_\chi$
(Sec.~\ref{sec:theory:band}) costs the bits nothing. The field's first mode carries its mean, which
the gradient echo reads, and takes 9 bits; the modes after it take 3 or 4 (Fig.~\ref{fig:bits}F).

\FloatBarrier
\subsection{The time-segment layout}\label{sec:theory:segments}

\begin{figure}[!tbp]
\centering
\includegraphics[width=\linewidth]{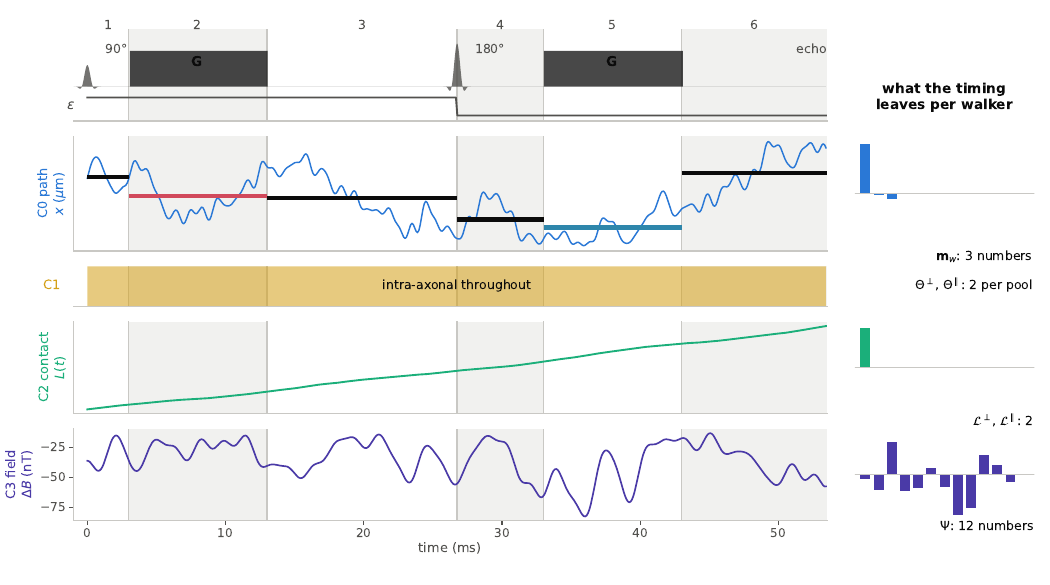}
\caption{Contract time, keep the knobs (\texttt{figures/fig\_segments.py}, data \texttt{fig\_segments\_data.py}).
One intra-axonal walker of the CACTUS pack under a spin echo with a pulse pair, $\delta=10$~ms,
$\Delta=30$~ms, $\TE=53.5$~ms, cut into six segments at the pulses' edges and the refocusing time. Its four series
as the pack decodes them: the path along the axon from its 256 bands, with each segment's mean position (bars; the
two under the pulses coloured); its pool; its cumulative local time; its field at 3~T, $B_0$ across the bundle. On
the right, what the timing contracts each to (each set scaled to its largest): the one-shape vector
$\mathbf m_w$ of Eq.~\eqref{eq:shape}, the gated pool times, the gated local times and the twelve stored gated
field integrals of Eq.~\eqref{eq:ch:susc}.}
\label{fig:segments}
\end{figure}

The pack keeps every time: any waveform and any gate within the band. A protocol uses few of them. It varies $b$
and direction far more often than its gate, the echo time, the refocusing times and the coherence pathway, and its
tissue and its field never touch time at all. So Eq.~\eqref{eq:channels} can be contracted along time alone, once per
timing, and left open along everything else (Fig.~\ref{fig:segments}). I call what that leaves the \emph{time-segment layout}; it is what a page stores. It is
exact, its only error the pack's truncation at $K$, and it gives up nothing but the waveform's course in time: a walker's 0.91~kB in the
pack become 37 bytes at one timing, while $b$, direction, pose, $T_1$, $T_2$, the relaxivities, the
susceptibilities and $B_0$ stay free (Table~\ref{tab:pack}).

\paragraph{C0, one shape, one vector.} A timing plays its gradient as a fixed temporal shape $\psi(t)$, the
refocusing signs included, at many amplitudes and directions, $\varepsilon(t)\,\mathbf G(t)=q\,\psi(t)\,\hat{\mathbf
g}$. Its phase is one vector per walker,
\begin{equation}
  \varphi_w=\scan{q\,\hat{\mathbf g}}\cdot\stored{\mathbf m_w},\qquad
  \stored{\mathbf m_w}=\tiss{\gamma}\int\scan{\psi(t)}\big(\stored{\rw(t)}-\stored{\rw(0)}\big)\,dt=\tiss{\gamma}\,\stored{\mathsf C_w}\scan{\boldsymbol\sigma},
  \label{eq:shape}
\end{equation}
with $\boldsymbol\sigma=\int\psi(t)\,\mathbf b(t)\,dt$ the shape's $K{+}2$ numbers, a projection of the stored bands,
its first entry, the start position, set to zero. Subtracting the start is exact, since a refocused waveform leaves
no net area, and keeps the stored values to displacements of micrometres. Nothing asks $\psi$ to be constant or
square: a pulse pair, a trapezoid with its ramps, a cosine oscillating gradient and a numerically designed waveform
are each one vector, exact to the pack's truncation at $K$. A measurement at that timing costs three multiplications
per walker instead of $3(K{+}2)$, 258 times fewer at $K=256$, and returns the same number to rounding for every
$b$-value and direction.

For a pulse pair, $\psi$ is $+1$ during the first pulse and $-1$ during the second, and
$\mathbf m_w=\tiss{\gamma}\delta\,(\bar{\mathbf r}_{w,1}-\bar{\mathbf r}_{w,2})$, with $\bar{\mathbf r}_{w,i}$ the
walker's time-averaged position during pulse $i$: the centre of mass of the path traced during it (the coloured bars
of Fig.~\ref{fig:segments}). This is the narrow-pulse phase with each position replaced by the walker's centre of
mass during its pulse \citep{MitraHalperin1995}; in a pore a long pulse puts that centre near the pore's middle,
which is why finite pulses make a pore look smaller. For any other shape, $\mathbf m_w$ is the path weighted by
$\psi$ in place of that equal-mass average.

\paragraph{C0, several shapes: segments and tensors.} A protocol that varies more than one amplitude within its
timing plays a sum of fixed shapes, $\varepsilon(t)\,\mathbf G(t)=\sum_k\psi_k(t)\,\mathbf g_k$, with the vectors
$\mathbf g_k$ set per measurement, and with the C1, C2 and C3 rows at the same gate a walker's exponent of
Eq.~\eqref{eq:channels} is
\begin{equation}
  \Lambda_w=\;i\sum_k\scan{\mathbf g_k}\cdot\stored{\mathbf m_{w,k}}
  \;-\;\sum_j\Big(\frac{\stored{\Theta^\perp_{w,j}}}{\tiss{T_{2,j}}}+\frac{\stored{\Theta^\parallel_{w,j}}}{\tiss{T_{1,j}}}\Big)
  \;-\;\frac{\tiss{\rho_2}\,\stored{\mathcal L^\perp_w}+\tiss{\rho_1}\,\stored{\mathcal L^\parallel_w}}{\tiss{D}}
  \;+\;i\,\varphi^{\chi}_w,
  \label{eq:segments}
\end{equation}
one vector $\stored{\mathbf m_{w,k}}=\tiss{\gamma}\,\stored{\mathsf C_w}\scan{\boldsymbol\sigma_k}$ per shape. Two families of shapes cover the
protocols a page serves. Time segments are the indicator shapes of the intervals between instants
$t_0<t_1<\dots<t_n$ that include every refocusing time and every switch of the waveforms to be served; their vectors
are $\gamma\tau_i\bar{\mathbf r}_{w,i}$, the segments' centres of mass times their durations and $\gamma$, and they serve every waveform
constant on the cuts, each segment at its own amplitude and direction. A waveform whose direction turns during the
encoding, as in $b$-tensor encoding \citep{Westin2016qti,Topgaard2017multidim}, is three shapes along fixed axes,
$\varepsilon(t)\,\mathbf G(t)=q\sum_{k=1}^{3}a_k\,\psi_k(t)\,\mathsf R\,\hat{\mathbf e}_k$, with $\mathsf R$ the
encoding's rotation and $a_k$ the weights of its shapes, and its phase is
\begin{equation}
  \varphi_w=\scan{q}\sum_{k=1}^{3}\scan{a_k}\,\big(\scan{\mathsf R}\,\hat{\mathbf e}_k\big)\cdot\stored{\mathbf m_{w,k}},
  \label{eq:tensor}
\end{equation}
three vectors per walker, nine numbers, which serve every rotation, every $b$-value and every weighting of the
three shapes, linear, planar or spherical. A rotation is a replay knob like a direction: it enters each walker's
phase, and the signal is the mean of the walkers' phase factors, so the encoding is exact in any substrate, not
only in the Gaussian limit in which a $b$-tensor alone describes it. Every set of shapes fixes the waveforms' course
in time and serves those waveforms alone; the gate, not the shapes, fixes the C1 to C3 integrals below, so shapes
that share a gate share them. The DiSCo page stores one shape at each of its pulsed and oscillating timings and the
three shapes of one tensor encoding (Sec.~\ref{sec:results:pages}).

\paragraph{C1 to C3, the gated integrals.} With the gate fixed the other three rows are integrals of the stored
series against it. Eq.~\eqref{eq:ch:c1} leaves two numbers per pool, the time the walker spent in pool $j$ while
transverse, $\Theta^\perp_{w,j}$, and while stored, $\Theta^\parallel_{w,j}$, and any $T_1$ and $T_2$ follow;
Eq.~\eqref{eq:ch:c2} leaves the gated local times $\mathcal L^\perp_w$ and $\mathcal L^\parallel_w$, and any
relaxivity follows; and C3 is the thirteen gated field integrals of Eq.~\eqref{eq:ch:susc}, from which every field
strength, susceptibility and field direction follows. A spin echo has no stored gate, so its walker carries one
occupancy time per pool and one local time.

\paragraph{The bits.} The contracted numbers are rounded by Eq.~\eqref{eq:container}, each with the weight of
what multiplies it: $S=\gamma^2G_{\max}^2\sum_i\tau_i^2$ for the segment means, over the segments the declared
envelope switches on; $S=q_{\max}^2$ for one shape's $\mathbf m_w$, the largest amplitude it declares for that
timing; $S=(\rho_{\max}/D)^2$ for the gated local times, which enter as a weight, $\exp(-\rho\mathcal L_w/D)$, with
the same two error terms; and $S=B_0^2\sum_cw_c^2$ for the gated field integrals, the field's weight of
Sec.~\ref{sec:theory:bits} with the gate already summed into them. Each inherits the rounding of the coefficients
it was contracted from, so the two shares add in quadrature and each takes $\epsilon/\sqrt2$.

\section{Methods}\label{sec:methods}

\FloatBarrier
\subsection{The replay pack}\label{sec:methods:pack}

A replay pack is one persistent simulation on one substrate as a file: the walker-side channels of
Eq.~\eqref{eq:channels} in the coefficient forms of Sec.~\ref{sec:theory:bridge}, truncated in modes by the rule of
Sec.~\ref{sec:theory:band}, and a header that says what the pack may be asked. Table~\ref{tab:pack} gives what that
costs, within the Monte-Carlo floor the certificate measures. Three
properties of the pack make a question cheap.

\paragraph{The pack is separable.} Every channel is its own tensor and every tier its own byte range, in one
\texttt{safetensors} file, a JSON header of named tensors followed by their bytes, memory-mapped rather than
parsed and readable from any language. A question therefore reads the tiers it needs and nothing else, and a
weaker scanner reads fewer modes of each (Table~\ref{tab:pack}). The same separability holds across walkers: a simulation too large for one file is
sharded over walkers in any grouping, and a consumer reads the walkers and the tiers of its question by byte range.

\paragraph{The pack is ordered.} Walker-indexed tensors are stored walkers first with the walkers shuffled and the
modes in band order, so the leading $n$ walkers are an unbiased sub-ensemble and the leading $K$ modes the
narrower band (Sec.~\ref{sec:theory:band}), both contiguous. In time a pack is a sequence of storage windows (100~ms
here), each certified on its own, with exact positions at every boundary (Eq.~\eqref{eq:bridge}): a pack is cut,
joined or continued at a window. Exact start positions let one simulation re-partition onto any voxel grid by
grouping walkers by where they started. Nothing is re-encoded to serve a smaller question.

\paragraph{The pack is quantised per band.} Each band is stored as an integer at its own scale, with the bits Eq.~\eqref{eq:container}
derives from the pack's floor target under a budget $\beta=0.1$ and $\kappa=1$, about the largest ratio
measured on the pores (0.4 to 1.1); where it reaches 1.1 the rounding errs a tenth more, well inside the
budget. At the 1\% floor of the DiSCo simulation that is 6.8 bits per band on average,
13 for the first eight bands and 5 or 6 for the last hundred, 653 bytes per walker for the positions. Table~\ref{tab:pack} gives what the
truncation and the quantisation each take off a walker; the certificate is measured on the quantised pack, so the
error it states includes the rounding.

The header carries what fixes the simulation and cannot be replayed (the fixed rows of Table~\ref{tab:knobs},
with the seed and the integrator settings), the envelope of Sec.~\ref{sec:theory:envelope}, and the certificate. The
envelope is stated and recorded, never a default: the scanner class with its per-axis $G_{\max}$ and $S_{\max}$,
from which the band $f_K$ of Eq.~\eqref{eq:band} follows (for the packs of this paper Connectome~2.0, 0.5~T/m and
600~T/m/s, so no largest $b$-value is declared); the largest surface relaxivity, $\rho_2\le3~\mu$m/s; the field
strengths with the largest susceptibility, $B_0$ of 1.5, 3 and 7~T with the substrate's own $\chi$; and the
refocusing depth the field channel serves. The certificate states the floor $f$, a standard error with the
rounding's share in it, beside each term of $\varepsilon_\mathrm{num}$ (Sec.~\ref{sec:theory:band}), so that
``solved'' is a checkable property of the file. The format is a published
specification \citep{FickRpkSpec2026}, and the datasets behind this paper, their revisions and their archive are
Sec.~\ref{sec:data}.

\begin{table}[h]
\centering\small
\caption{What DiSCo's walk costs to keep, over DiSCo's 53.5~ms and over the 100~ms the packs are built for; a pack
answers every echo time beneath its own (Sec.~\ref{sec:theory:envelope}). The first two rows are DiSCo's walk as
simulated \citep{RafaelPatino2021disco}, $10^5$ steps of 0.535~$\mu$s, positions only, in float32: at its own
$10^9$ walkers, a standard error of at most 0.8\%, and at the $3.21\times10^8$ walkers of a 1\% floor. Every other row is
at that 1\% floor, the DiSCo pack's. \emph{Trajectory}: float32 at every step of the walk, 3\,015 steps of
33~$\mu$s over 100~ms, the position and, for every tier, the field's twelve channels and the contact.
\emph{Truncated}: the band of Sec.~\ref{sec:theory:band}, 256 modes over 100~ms, still float32. \emph{Pack}: the
bands quantised in the container derived at the 1\% floor (Sec.~\ref{sec:methods:pack}); at 53.5~ms the same
container on the leading 137 bands. \emph{Prisma}: the same bytes, read to the leading 104 of the 256 modes.
\emph{Time-segment layout}: Eqs.~\eqref{eq:shape} and \eqref{eq:ch:susc} contracted at stored timings and gates,
every other knob free: $b$, direction, pose, $T_1$, $T_2$, $\rho_2$, $\chi$, $B_0$ and its direction. Ratios are
against the trajectory at the same floor. $^*$Published: \href{https://huggingface.co/datasets/SubstrateCommons/disco-replay}{\texttt{SubstrateCommons/disco-replay}}
\citep{DiscoReplayHub2026} and \href{https://huggingface.co/datasets/SubstrateCommons/disco-time-segments}{\texttt{SubstrateCommons/disco-time-segments}}
\citep{DiscoTimeSegmentsHub2026}, the layout the page reads.}
\label{tab:pack}
\begin{tabular}{@{}lrrrrrr@{}}
\toprule
& \multicolumn{3}{c}{Every tier} & \multicolumn{3}{c}{Diffusion only} \\
\cmidrule(lr){2-4}\cmidrule(lr){5-7}
& 53.5~ms & 100~ms & smaller & 53.5~ms & 100~ms & smaller \\
\midrule
\textit{DiSCo as simulated, $10^9$ walkers (0.8\%)}      & -- & -- & -- & \textit{1.2~PB} & -- & -- \\
\textit{DiSCo as simulated, at a 1\% floor}               & -- & -- & -- & \textit{385~TB} & -- & -- \\
\midrule
Trajectory, every step of the walk            & 33~TB   & 62~TB   & $1\times$      & 6.2~TB  & 12~TB   & $1\times$ \\
Truncated, Connectome~2.0 band, float32       & 0.80~TB & 1.5~TB  & $41\times$     & 0.55~TB & 1.0~TB  & $12\times$ \\
Replay pack, the same bands quantised         & 177~GB  & 294~GB$^*$ & $210\times$ & 138~GB  & 220~GB  & $53\times$ \\
\quad read by a Prisma                        & 111~GB  & 186~GB  & $330\times$    & 70~GB   & 112~GB  & $100\times$ \\
Time-segment layout, one stored timing        & 12~GB   & 12~GB   & $5\,200\times$ & 5.1~GB  & 5.1~GB  & $2\,300\times$ \\
Time-segment layout, the page's nine timings  & 51~GB$^*$ & 51~GB$^*$ & $1\,200\times$ & -- & -- & -- \\
\bottomrule
\end{tabular}
\end{table}

\FloatBarrier
\subsection{Encoding, replaying and sharing replay packs}\label{sec:methods:engine}

\paragraph{Encoding: \dmipysim{}.} Packs are written by \dmipysim{}\footnote{\url{https://github.com/dmrai-lab/dmipy-sim}, on PyPI as \texttt{dmipy-sim}; AGPL-3.0.},
a Monte Carlo engine written for persistence and scale, with the multiphysics content of \citet{Cottaar2026mcmr},
every row of Eq.~\eqref{eq:channels} active in one substrate at once, on analytic pores and their packings, curved
and myelinated tubes, sphere unions and triangle meshes \citep{HallAlexander2009camino,RafaelPatino2020mcdc,Cottaar2026mcmr}.
It is written in JAX \citep{Bradbury2018jax}, which compiles one program to the CPU or any GPU. The walk
accumulates the series of Sec.~\ref{sec:theory:bridge} as coefficients on the fly, so its memory stays flat in the
walk's length and no trajectory is written. Its step takes one of two lengths chosen to match the Gaussian step's
moments \citep{KloedenPlaten1992sde}, at the largest duration that holds the walk's diffusion spectrum within
$\epsilon_D=10^{-3}$ up to the acquisition class's band, measured from two pilot walks and recorded in the pack. Every pack carries its certificate, the terms of Sec.~\ref{sec:theory:band} and the
Monte-Carlo floor, in its header.

\paragraph{Replaying: \texttt{dmipy-replay}.} Packs are replayed by \texttt{dmipy-replay}\footnote{\url{https://github.com/dmrai-lab/dmipy-replay}, on PyPI as \texttt{dmipy-replay}; Apache-2.0.},
a client that depends on NumPy alone: it fetches a pack at a pinned revision, checks its sha256, reads its tiers by
range, reads Pulseq sequences \citep{Layton2017pulseq} natively, and replays them bit for bit as \dmipysim{} does,
which imports it. A set of $\Nq$ measurements is one product of the $\Nw\times3(K{+}2)$ coefficients with a compiled
$3(K{+}2)\times\Nq$ acquisition, Sec.~\ref{sec:theory:bridge}. A third package, \texttt{dmipy-design}\footnote{\url{https://github.com/dmrai-lab/dmipy-design}, on PyPI as \texttt{dmipy-design}; AGPL-3.0.},
reads packs through the same description of an acquisition and optimises waveforms under a scanner's limits; its
search for the worst waveform a scanner class can play gives the waveforms the band certificate is tested on
(Sec.~\ref{sec:results:band}).

\paragraph{Sharing: SubstrateCommons.} Packs are shared from the \texttt{SubstrateCommons} organisation on the
Hugging Face Hub\footnote{\url{https://huggingface.co/SubstrateCommons}}, one dataset per substrate family. Each
dataset holds its packs, a manifest with every pack's sha256, size, walk parameters, certificate and licences, and
a substrate card rendered from the manifest. A pack is addressed by dataset, path and revision, and a reader checks
it against the manifest at that revision, so a figure cites an immutable object. The Hub serves byte ranges, and
since every channel, tier and window of a pack is its own byte range (Sec.~\ref{sec:methods:pack}), a question
fetches only the walkers, modes and tiers it needs. Every pack is under CC BY-NC-SA 4.0 or a commercial licence,
stated per pack in the manifest beside the licence of its source geometry.

\FloatBarrier

\subsection{The experiments}\label{sec:methods:experiments}

\begin{figure}[!htbp]
\centering
\includegraphics[width=\linewidth]{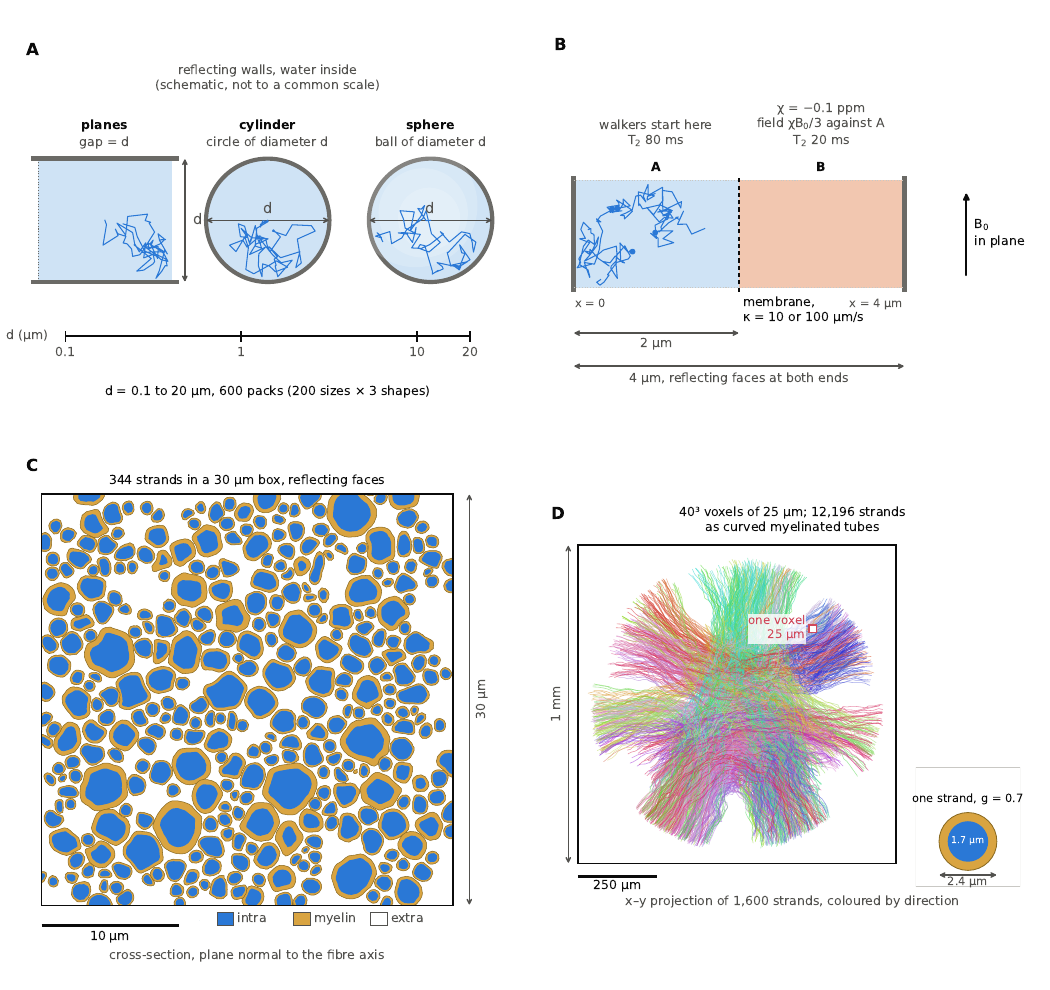}
\caption{The four substrates of the experiments (\texttt{figures/fig\_substrates.py}). (A) Canonical pores:
parallel planes at gap $d$, a cylinder and a sphere of diameter $d$, reflecting walls and water inside, $d$ from
0.1 to 20~$\mu$m in 200 sizes per shape, 600 replay packs; the shapes are drawn at one size and the axis gives
$d$. (B) The permeable slab: 4~$\mu$m with reflecting faces and a membrane at 2~$\mu$m of permeability 10 or
100~$\mu$m/s; walkers start in compartment A ($T_2$ 80~ms), and compartment B ($T_2$ 20~ms) is magnetised,
$\chi=-0.1$~ppm, a field of $\chi B_0/3$ against A with $B_0$ in the plane of the faces. (C) The CACTUS bundle
\citep{VillarrealHaro2023}: a section normal to the fibre axis of the 344 strands in their 30~$\mu$m box with
reflecting faces, from the meshes the pack is walked on; intra-axonal, myelin and extra-axonal pools. (D) DiSCo
\citep{RafaelPatino2021disco}: 1\,600 of its 12\,196 strands, projected and coloured by direction, in the 1~mm
cube of $40^3$ voxels of 25~$\mu$m (one outlined); each strand is two tubes along its centreline, the axolemma at
the listed diameter (median 1.7~$\mu$m) and a sheath at a $g$-ratio of 0.7 (inset).}
\label{fig:substrates}
\end{figure}

Table~\ref{tab:experiments} lists every experiment of Sec.~\ref{sec:results}: what is checked, on what substrate
and simulation, under which acquisition, against which reference; Fig.~\ref{fig:substrates} shows the four substrates. The paragraphs below give the one design
decision of each that a reader needs to judge its result; the recipes, sources and solver settings are in
the Supplementary Material, S1, and every figure names the script that produced it.

\begin{table}[!htbp]
\centering\footnotesize\setlength{\tabcolsep}{4pt}\renewcommand{\arraystretch}{1.1}
\caption{The experiments. Simulations at $D=2\times10^{-9}$~m$^2$/s unless stated, stored at the storage band of
Sec.~\ref{sec:results:band} (256 modes per 100~ms); ``floor'' is the split-half Monte-Carlo floor.}
\label{tab:experiments}
\begin{tabular}{@{}p{0.16\linewidth}p{0.27\linewidth}p{0.22\linewidth}p{0.25\linewidth}l@{}}
\toprule
Check & Substrate, simulation & Acquisition & Reference & Fig.; Supp. \\
\midrule
Band certificate & free water, 2 and 10~$\mu$m cylinders, CACTUS bundle; 20\,000 walkers, 100~ms, 2\,000 modes & the worst deliverable waveform of six scanner classes, PGSE, OGSE & the truncation theorem, Eq.~\eqref{eq:tail}; split-half & \ref{fig:band}; \ref{app:band} \\[2pt]
Canonical pores & the published family \citep{CanonicalPoresHub2026}: planes, cylinders, spheres, 0.1 to 20~$\mu$m in steps of 0.1, 600 packs; walkers set per pack by its pilot for a standard error of $5\times10^{-3}$, 8\,000 to 40\,000; 100~ms & four PGSE timings at $b$ 1000, 3000; OGSE 50 to 800~Hz; FID & matrix method; Brownstein--Tarr; both under Robin & \ref{fig:pores}; \ref{app:pores} \\[2pt]
Exchange & permeable slab, 4~$\mu$m, $\kappa$ 10 and 100~$\mu$m/s; 40\,000 walkers, 100~ms & FID, spin echo, PGSE, gradient echo, TE 10 to 100~ms, 3 and 7~T & 1-D Bloch--Torrey with the transmission condition & \ref{fig:slab}; \ref{app:slab} \\[2pt]
Realistic axons & CACTUS bundle (111\,011 walkers, $D=1.7\times10^{-9}$); 100~ms & PGSE, OGSE 25 to 400~Hz, spin echo, gradient echo at 3 and 7~T, relaxivity 0 to 3~$\mu$m/s & the pack's own floor; MCMRSimulator.jl and MC/DC strand by strand, as packs & \ref{fig:mesh}; \ref{app:meshes} \\[2pt]
DiSCo, partitioned & 12\,196 strands in 1~mm$^3$, as curved myelinated cylinders on its centrelines, $D=0.6\times10^{-9}$; $3.2\times10^8$ walkers on the 25~$\mu$m grid & the dataset's four shells; five schemes it was never scanned with; a $b$-tensor encoding & the published images; the known connectome & \ref{fig:disco}; \ref{app:disco} \\
\bottomrule
\end{tabular}
\end{table}

\paragraph{The band certificate.}\label{sec:methods:band}
The substrates are simulated once and kept as float coefficients to 2\,000 modes over 100~ms
\citep{CoefficientWalksHub2026}, so every $K$ is a leading slice of one simulation. Per substrate, scanner class (Table~\ref{tab:classes}) and $K$, the certificate is the largest $|S_\mathrm{full}-S_K|$ over waveforms within the
class's amplitude and slew limits that vanish at both ends and refocus, found by projected ascent on half the walkers and its error measured on the other half against Eq.~\eqref{eq:tail}, which
for free water needs no simulation at all.

\paragraph{Canonical pores.}\label{sec:methods:canonical}
Each pack's walker count is set by its own pilot for a
standard error of $5\times10^{-3}$ under any toggle, $n = V/f^2$ with the per-walker variance $V\le1$, so a sub-micron
pore has 8\,000 walkers and the 525 largest 40\,000; every comparison is a pore's worst row against the level its
walkers' noise alone reaches in 99 draws of 100: the gradient channel against the matrix method, the surface channel against
Brownstein--Tarr, the two together under the Robin condition, each reference stepped finely enough to resolve the
pore's own relaxation.

\paragraph{The compartment and field channels.}\label{sec:methods:c1c3}
The two channels the pores do not exercise are checked on the one substrate where their physics has an exact
solution, a slab split by a membrane with compartment $T_2$ of 80 and 20~ms and one compartment magnetised, a
layer whose field is $\chi B_0/3$ against the other with $B_0$ in its plane. Every walker starts in the
unmagnetised compartment, so the other fills through the membrane at a rate its permeability sets, about
200~ms at 10~$\mu$m/s and 20~ms at 100~$\mu$m/s: the compartment channel must place each crossing in time and the
field channel follow phase in a compartment whose population changes, both against the same initial-value
problem solved exactly.

\paragraph{Realistic axons.}\label{sec:methods:meshes}
The CACTUS bundle \citep{VillarrealHaro2023} is simulated once with every channel of
Eq.~\eqref{eq:channels}, its field channel at 128 modes. Its references are two other engines,
MCMRSimulator.jl \citep{Cottaar2026mcmr} and MC/DC \citep{RafaelPatino2020mcdc}, each run strand by strand on
the same meshes from the same start positions, their walks stored as packs and read by the same reader.

\paragraph{The replay phantom.}\label{sec:methods:phantoms}
DiSCo \citep{RafaelPatino2021disco} is simulated from its published centrelines rather than its meshes, at its
diffusivity: each strand is a curved myelinated cylinder, the tube a ball sweeps along the centreline at the listed
diameter with a sheath around it, a sphere-swept line of collision detection \citep{Larsen2000ssv}, cylindrical along every segment and spherical at every joint. A wall is then a
distance to a segment, found through a grid over the segments, and a step is reflected in closed form: one test per
nearby segment where a twelve-sided mesh of the same tube has at least 24 triangles, so the 636\,951 segments stand
in for some 15 million triangles per wall. The sheath holds no water but is a susceptibility source, so one simulation
carries all four channels in both water pools, and the dataset's own simulation is the replay with the field
off. The susceptibility field must reach the
whole millimetre, since stopping its sum at 50~$\mu$m still moves it by 8.1\%. Each segment
contributes the dipole kernel integrated in closed form along it; the segments near a walker are summed exactly
where it stands, and the rest through a grid built once per substrate and shipped with it, the particle--mesh split
of \citet{HockneyEastwood1988} that cosmological $N$-body codes use for gravity \citep{Efstathiou1985}, here with
sources that never move. An independent $k$-space route that shares no closed
form with this sum agrees with it to 0.15\% (Appendix~\ref{app:field}). The walkers are planned for the most a 300~GB dataset on the Hub holds: from
each voxel's water shares and a pilot's per-walker cost at the class envelope, the plan gives every voxel and pool
the walkers that bring each voxel to one floor, a standard error of 1.0\% that holds under any combination of
tiers, and so places $3.21\times10^8$ walkers over the 31\,815 voxels of its 25~$\mu$m grid that hold water,
assigned to voxels by where they start. The walk is one pack of 294~GB (Sec.~\ref{sec:results:pages}).
The page's layout holds each walker's vectors as integers at the precision Sec.~\ref{sec:theory:segments} derives
for each timing's largest $q$, within the timing's rounding budget, decoded in the kernel.
The acquisitions it was never scanned with are timing classes of the page's layout: a clinical shell, a
three-shell research scheme, a Connectome~2.0 shell, a cosine OGSE at 50~Hz, a stimulated echo (PGSTE), and linear and spherical $b$-tensor
encoding at $b$ of 100 to 2000~s/mm$^2$ from the three gradient shapes of one numerically optimised spherical
waveform \citep{Sjolund2015now}; their timings and the tractography pipeline are Supplement~\ref{app:disco}.

\section{Results}\label{sec:results}

\FloatBarrier
\subsection{The modes a pack needs are set by the scanner}\label{sec:results:band}

\begin{figure}[!htbp]
\centering
\includegraphics[width=\linewidth]{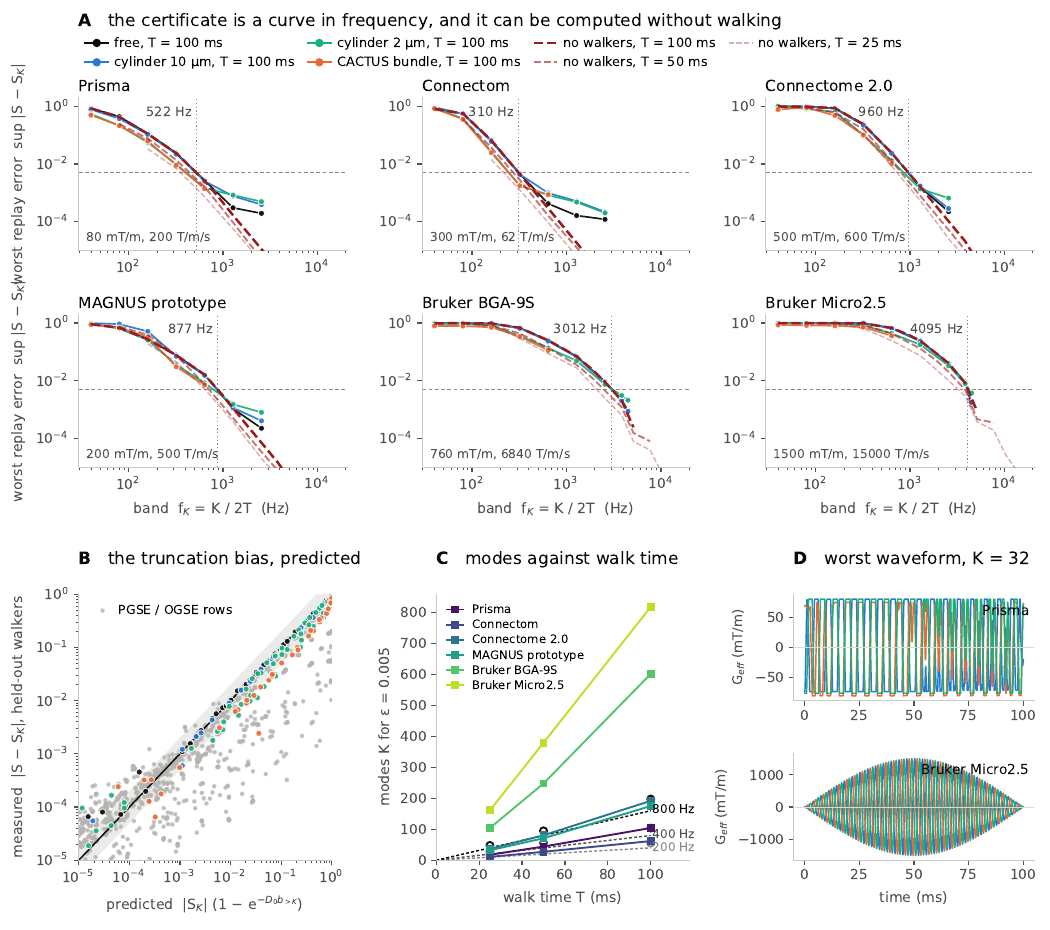}
\caption{A pack is a band (\texttt{figures/fig\_band.py}, data \texttt{fig\_band\_data.py}).
\textbf{A}: per scanner class, the worst replay error over the deliverable waveforms against the
band $f_K=K/2T$: found on the walkers of four 100~ms simulations kept as float coefficients \citep{CoefficientWalksHub2026} (points; the CACTUS bundle to its stored 256 modes, 640~Hz), and found on
Eq.~\eqref{eq:tail} for free diffusion with no simulation (red dashed, for 100, 50 and 25~ms). The
grey dashed line is $\varepsilon=5\times10^{-3}$; the dotted line its walker-free crossing.
\textbf{B}: the truncation bias on held-out walkers against Eq.~\eqref{eq:tail} from the
waveform alone, for every worst waveform with $|S_K|\ge0.05$ (colour) and for PGSE and cosine-OGSE
rows (grey); the band is a factor of two. \textbf{C}: modes for $\varepsilon$ against simulation
time: the walker-free certificate per class (squares), and the crossing of three OGSE
frequencies on the Micro2.5 (circles) with their lines $K=2fT$. \textbf{D}: the worst waveform the search finds for a pack cut at $K=32$ (160~Hz), walker-free, for the weakest
and the strongest class.}
\label{fig:band}
\end{figure}

The scanner, not the tissue, sets how many modes a pack keeps (Sec.~\ref{sec:methods:band}).
Table~\ref{tab:classes} lists six gradient systems, from a clinical 3~T scanner to a microscopy insert, by their
per-axis amplitude and slew rate, and the band a pack needs for each at a tolerance of $5\times10^{-3}$
(Fig.~\ref{fig:band}A). The band follows the slew rate, not the amplitude: the Connectom, with almost four times the
Prisma's amplitude but a third of its slew rate, needs a narrower band, because the highest frequency at which a
gradient can still deposit diffusion weighting is set by how fast it can switch. These bands come from a closed form
in the waveform alone, Eq.~\eqref{eq:tail}, computed without a simulation, which bounds the replay error of every
waveform the scanner can play: the bound lies above every measured point that clears the walkers' own floor, and its
worst waveform, replayed on the free-water simulation, agrees with it to two percent.

\begin{table}[!htbp]
\centering\small
\caption{The six gradient systems of Fig.~\ref{fig:band}: per-axis amplitude and slew rate, and the band a pack
needs for every waveform the system can play at a tolerance of $5\times10^{-3}$, from Eq.~\eqref{eq:tail}. Sources
in Supplement Table~\ref{tab:scanners}.}
\label{tab:classes}
\begin{tabular}{@{}llrrr@{}}
\toprule
Class & System & $G_{\max}$ (mT/m) & $S_{\max}$ (T/m/s) & Band (Hz) \\
\midrule
Prisma & clinical 3~T scanner & 80 & 200 & 522 \\
Connectom & human connectome scanner, derated for diffusion & 300 & 62.5 & 310 \\
MAGNUS & high-performance head insert & 200 & 500 & 877 \\
Connectome~2.0 & next-generation connectome scanner & 500 & 600 & 960 \\
BGA-9S & preclinical, 9~cm bore & 760 & 6\,840 & 3\,000 \\
Micro2.5 & microscopy insert, 4~cm bore & 1\,500 & 15\,000 & 4\,100 \\
\bottomrule
\end{tabular}
\end{table} The restricted substrates fall below
the free-water curve, as they must, since a wall can only lower a path's diffusion spectrum, never raise it; the few
points above it, by at most 1.4 times, are the search's noise on 20\,000 walkers. Fig.~\ref{fig:band}D shows what
sets those bands: for a pack cut at $K=32$ (160~Hz), the waveform that loses the most signal, on the Prisma a train
of the fastest lobes its slew rate allows, on the Micro2.5 an oscillation far above the band.

Fig.~\ref{fig:band}B asks more of Eq.~\eqref{eq:tail}: not a bound but a prediction of the error of one waveform,
measured on the walkers the search did not use against $|S_K|(1-e^{-D_0b_{>K}})$ computed from the waveform alone. The median ratio of measured to predicted error is 1.00
on free water, 0.91 and 0.97 on the 2 and 10~$\mu$m cylinders, 0.55 on the CACTUS bundle, whose pools restrict the
path most, and 1.05 on ordinary PGSE and OGSE rows, over errors from $10^{-6}$ to the whole signal.

Fig.~\ref{fig:band}C separates two questions Eq.~\eqref{eq:band} does not. For one declared acquisition, the modes a
longer simulation needs grow linearly, $K=2fT$ to within a factor of 1.2 to 1.8 set by the ramps. For the certificate over
everything a scanner can play they grow faster, as $T^{1.17}$ to $T^{1.28}$, because a longer storage window lets a waveform
put more $b$ above the band at the same frequency.

This fixes the storage of everything that follows. The packs of Secs.~\ref{sec:results:canonical}
to~\ref{sec:results:phantoms} are simulated for 100~ms and stored at $K=256$, a band of 1.28~kHz that covers every
human scanner of Table~\ref{tab:classes}, Connectome~2.0 with a margin of 1.33 and the others with more, of which a
Prisma reads the leading 104 modes. A longer simulation declares a gradient-on budget of 100~ms
and keeps the same bandwidth, so its modes grow linearly with its length, and a shorter echo time is a prefix at the
same number of bands per second.

\FloatBarrier
\subsection{Canonical pores agree with the closed forms}\label{sec:results:canonical}

\begin{figure}[!htbp]
\centering
\includegraphics[width=\linewidth]{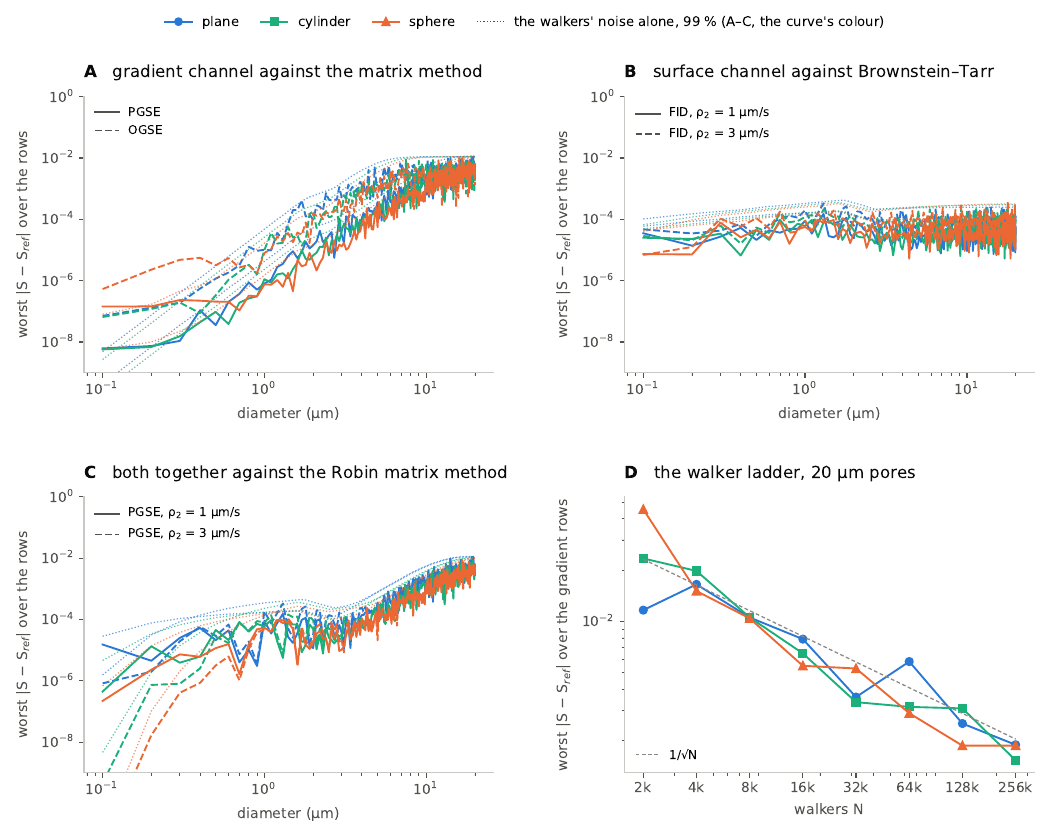}
\caption{The published family of canonical pores against the closed forms (Sec.~\ref{sec:methods:canonical}; 600
packs, 8\,000 to 40\,000 walkers each): the worst residual over a pore's rows against its diameter, with, dotted, the
level that worst row reaches from the walkers' noise alone in 99 draws of 100. Colour is the pore's shape in every
panel; solid and dashed lines are the two series each panel names. (A) The gradient channel, PGSE and
OGSE, against the matrix method. (B) The surface channel, free-induction decay at two relaxivities, against
Brownstein--Tarr. (C) Both, PGSE with relaxivity, against the Robin matrix method. (D) The walker ladder on the
20~$\mu$m pores walked again at 256\,000 walkers: the worst gradient-row residual over nested prefixes of the shuffled
ensemble against walker count, with the $1/\sqrt N$ line.}
\label{fig:pores}
\end{figure}

Figure~\ref{fig:pores} puts the 600 canonical pores \citep{CanonicalPoresHub2026} against their closed forms
(Sec.~\ref{sec:methods:canonical}). On the gradient channel (A) the residual follows that level over six decades of signal: of the 6\,300 PGSE and OGSE
rows above one micron the largest is 4.2 standard errors, the extreme of that many draws, and the one exception is
the cosine OGSE at 800~Hz, the battery's fastest waveform, whose residual of $2\times10^{-5}$ is the pack's
truncation at 256 modes, inside the $2.3\times10^{-4}$ its certificate states. Below one micron the residual stays
under $1.3\times10^{-5}$. On the surface channel (B) the free-induction decay lies within $3.8\times10^{-4}$ of
Brownstein--Tarr at both relaxivities and every diameter, at the noise level. The two channels together (C), PGSE with relaxivity
against the Robin matrix method, follow the noise level as the gradient channel does. The walker ladder (D), the 20~$\mu$m pores
walked again to 256\,000 walkers, shows the worst gradient-row residual falling as $1/\sqrt N$, from
$1$--$4.5\times10^{-2}$ at 2\,000 walkers to $1.5$--$1.9\times10^{-3}$, with no floor in sight.

\FloatBarrier
\subsection{The compartment and field channels replay exchange}\label{sec:results:c1c3}

\begin{figure}[!htbp]
\centering
\includegraphics[width=\linewidth]{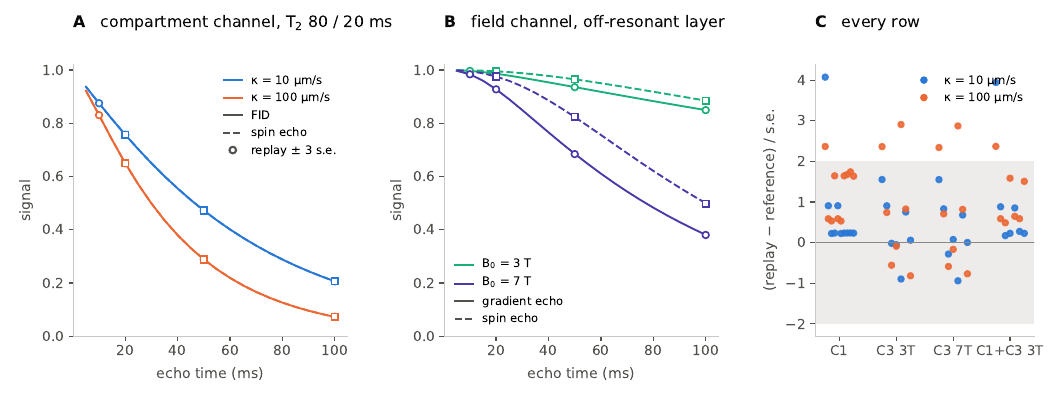}
\caption{The compartment and field channels against the one-dimensional Bloch--Torrey reference on a permeable
slab (Sec.~\ref{sec:methods:c1c3}). (A) The compartment channel: signal against echo time for two pools with $T_2$
of 80 and 20~ms at two permeabilities, lines the reference (FID solid, spin echo dashed), markers the replay with
three standard errors. (B) The field channel: the same with the magnetised compartment at 3 and 7~T, gradient and spin
echo. (C) Every row's residual in units of its standard error.}
\label{fig:slab}
\end{figure}

Figure~\ref{fig:slab} tests the compartment channel, where each walker is and for how long, and the field channel,
the susceptibility field each walker passes through, against the exact one-dimensional solution on the permeable
slab of Fig.~\ref{fig:substrates}B (Sec.~\ref{sec:methods:c1c3}). One simulation of the slab answers four
acquisitions at echo times from 10 to 100~ms: the free-induction decay; the spin echo, which refocuses the static
field and has no diffusion gradient; the PGSE, the same spin echo with diffusion gradients; and the gradient echo,
which keeps the field's dephasing. Each is replayed at both membrane permeabilities, 10 and 100~$\mu$m/s, and at
3 and 7~T, 62 signal values in all. None differs from the exact solution by more than 4.1 times its Monte Carlo
standard error, and on average by 0.8 times it (panel C); in absolute terms the replay is within $1.0\times10^{-3}$
of the exact signal for the compartment channel and $2.0\times10^{-3}$ for the field channel, the largest at the
100~ms spin echo at 7~T, where the magnetised compartment dephases most. The spin echo refocuses the static part of
that compartment's field, and what it leaves is the dephasing of walkers that crossed the membrane during the echo
(panel B): a model that gives each compartment its own off-resonance cannot produce that term, and the field channel
replays it because it stores the field along each walker's path rather than per compartment.

\FloatBarrier
\subsection{Realistic axons: three engines agree}\label{sec:results:meshes}

\begin{figure}[!htbp]
\centering
\includegraphics[width=\linewidth]{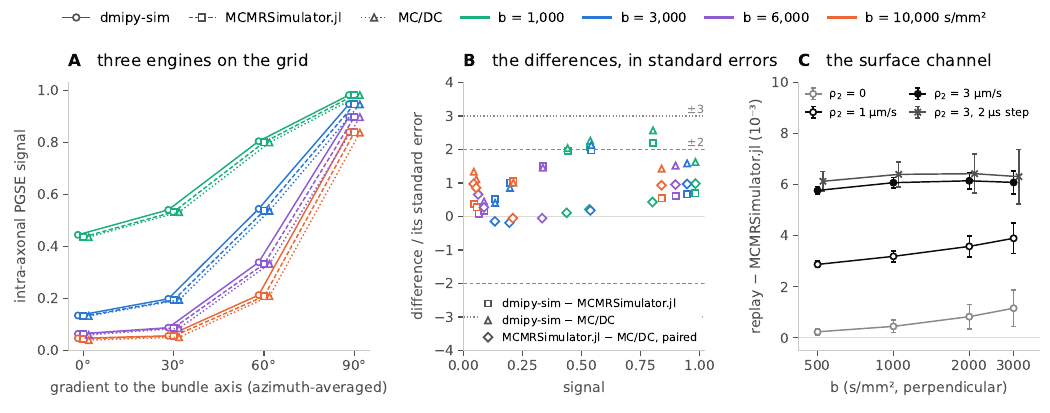}
\caption{The CACTUS bundle's intra-axonal pool walked by three engines, each walk a replay pack read by one reader
(\texttt{figures/fig\_mesh.py}; Sec.~\ref{sec:methods:meshes}). (A) PGSE signal ($\delta/\Delta=10/40$~ms) against
the gradient's angle to the bundle axis, averaged over 11 azimuths, at four $b$-values; bars one standard error.
(B) Every reading's difference between two engines in units of its standard error, against the signal, with
$\pm2$ and $\pm3$. (C) The surface channel: this engine's replay minus MCMRSimulator against $b$, across the
bundle, at $\rho_2$ of 0, 1 and 3~$\mu$m/s, and at 3~$\mu$m/s with MCMRSimulator at a 2~$\mu$s step.}
\label{fig:mesh}
\end{figure}

On a realistic substrate there is no closed form, so the reference is other simulators. The intra-axonal pool of the
CACTUS bundle's 176 strands was walked by three engines, this one, MCMRSimulator.jl \citep{Cottaar2026mcmr} and
MC/DC \citep{RafaelPatino2020mcdc} (Sec.~\ref{sec:methods:meshes}), each walk stored as a replay pack and all three
read by one reader. Read under PGSE at $b$ of 1\,000 to 10\,000~s/mm$^2$ with the gradient from along the bundle to
across it, a grid on which the signal runs from 0.04 to 0.98, the three agree within three standard errors on every
one of the 16 readings (Fig.~\ref{fig:mesh}A, B; Table~\ref{tab:enginesgrid}). MCMRSimulator and MC/DC, which walk
the same start positions, agree within one standard error; this engine reads above both, by up to 2.6 standard
errors, most along the bundle at $b=1\,000$~s/mm$^2$, and its pack walls the bundle's box with reflecting faces where
the other two continue it periodically (Supplement~\ref{app:meshes}). On the surface channel this engine's contact
matches Brownstein--Tarr on tessellated cylinders (Supplement~\ref{app:meshes}), and on the strands the replay sits
above MCMRSimulator by 0.3\% of the signal at $\rho_2=1$~$\mu$m/s and 0.6\% at 3~$\mu$m/s, at either of
MCMRSimulator's steps (Fig.~\ref{fig:mesh}C). The other two engines' walks took 1.8 and 1.4~hours on eight CPU
threads; a row of any of the three packs replays in 50 to 150~ms on one CPU core.

\begin{table}[!htbp]
\centering\footnotesize
\caption{The three engines on the CACTUS intra-axonal pool: differences of the replayed PGSE signal
($\delta/\Delta=10/40$~ms, averaged over 11 azimuths around the bundle axis) along the bundle ($0^\circ$) and across
it ($90^\circ$), as difference $\pm$ standard error ($z$), in units of $10^{-3}$; this engine's columns as
independent walks, the last paired on the two engines' shared start positions.}
\label{tab:enginesgrid}
\begin{tabular}{rr ccc}
\toprule
$b$ (s/mm$^2$) & angle & dmipy-sim $-$ MCMRSimulator.jl & dmipy-sim $-$ MC/DC & MCMRSimulator.jl $-$ MC/DC \\
\midrule
1,000 & 0$^\circ$ & $+8.3\pm4.3$ (+2.0) & $+8.7\pm4.3$ (+2.0) & $+0.5\pm4.2$ (+0.1) \\
1,000 & 90$^\circ$ & $+0.1\pm0.2$ (+0.7) & $+0.3\pm0.2$ (+1.6) & $+0.2\pm0.2$ (+1.0) \\
3,000 & 0$^\circ$ & $+2.6\pm5.1$ (+0.5) & $+2.1\pm5.1$ (+0.4) & $-0.7\pm5.0$ (-0.1) \\
3,000 & 90$^\circ$ & $+0.4\pm0.5$ (+0.7) & $+0.9\pm0.5$ (+1.6) & $+0.5\pm0.5$ (+1.0) \\
6,000 & 0$^\circ$ & $+0.4\pm5.1$ (+0.1) & $+3.7\pm5.1$ (+0.7) & $+3.3\pm5.0$ (+0.7) \\
6,000 & 90$^\circ$ & $+0.6\pm1.0$ (+0.6) & $+1.5\pm1.0$ (+1.5) & $+0.9\pm1.0$ (+1.0) \\
10,000 & 0$^\circ$ & $+1.9\pm5.1$ (+0.4) & $+6.8\pm5.1$ (+1.3) & $+4.9\pm5.0$ (+1.0) \\
10,000 & 90$^\circ$ & $+0.8\pm1.5$ (+0.5) & $+2.2\pm1.5$ (+1.4) & $+1.4\pm1.5$ (+0.9) \\
\bottomrule
\end{tabular}
\end{table}

\FloatBarrier
\subsection{DiSCo, a partitioned replay phantom}\label{sec:results:phantoms}
\begin{figure}[!htbp]
\centering
\includegraphics[width=\linewidth,height=0.72\textheight,keepaspectratio]{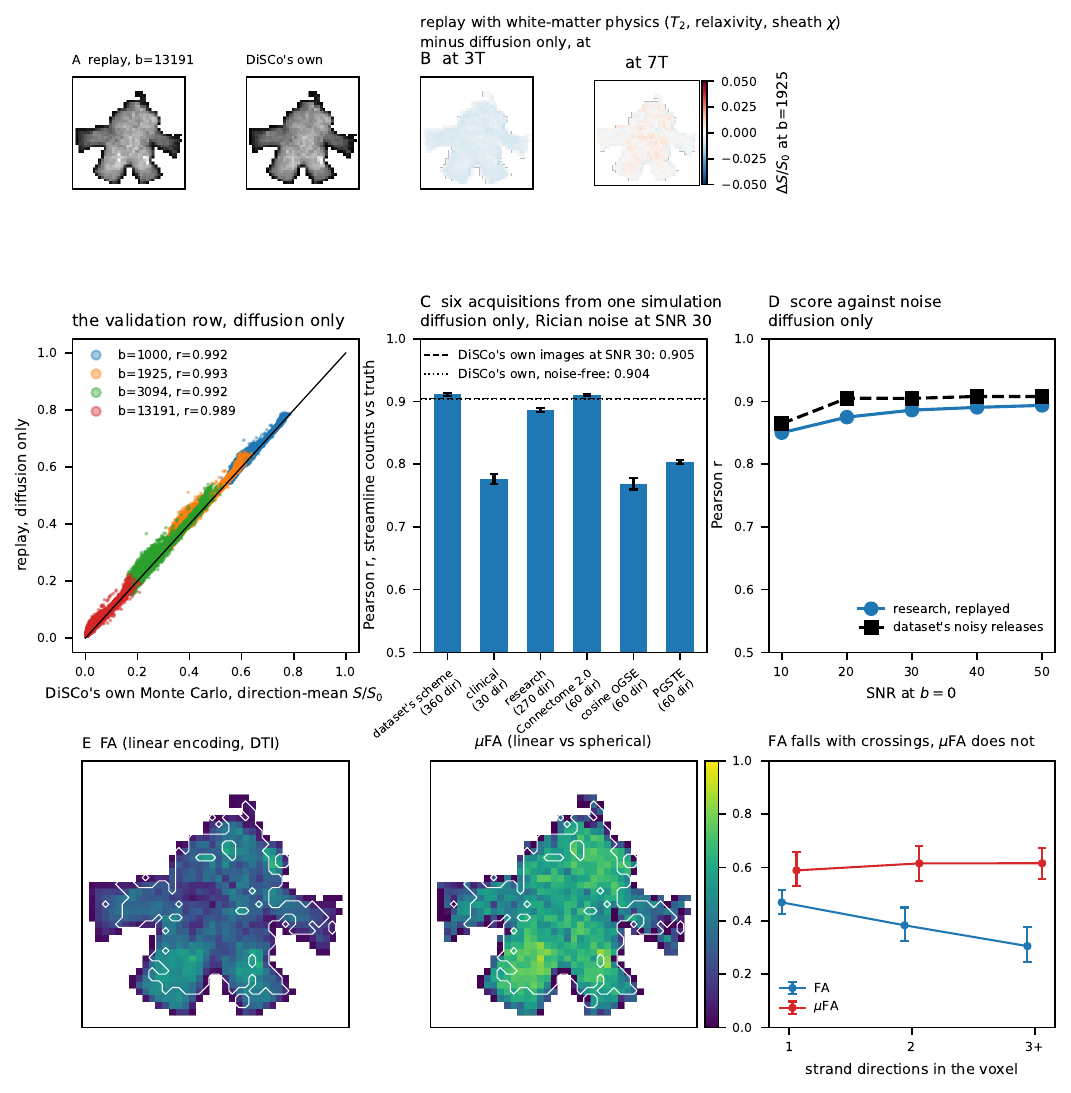}
\caption{DiSCo, partitioned: one simulation, the dataset's protocol and five others. (A) The dataset's own
protocol replayed from the pack against its published noise-free images: one slice at $13\,191$~s/mm$^2$,
and the direction-mean signal of every mask voxel per shell (the scatter is the pass's floor). (B) The same protocol with the catalogue's white matter at 3 and 7~T, as the
change of the $b=1925$ shell's direction mean. (C) The tractography score of six acquisitions replayed
from the simulation at SNR 30 (three seeds; bars are the spread), with the dataset's own images through the
same pipeline as the ceiling. (D) The research scheme's score against SNR beside the dataset's own
releases. (E) $b$-tensor encoding: fractional anisotropy (DTI on the linear encoding) and microscopic anisotropy
on one slice, crossing voxels outlined from the ground truth, and both against the number of strand directions in
the densest quarter of strand voxels.}
\label{fig:disco}
\end{figure}

Fig.~\ref{fig:disco}A puts the dataset's own protocol \citep{RafaelPatino2021disco}, replayed from the pack,
voxel by voxel against its published noise-free images over the 15\,040 voxels of its mask: they agree to a median $|\Delta S/S_0|$ of 0.006 to 0.010 on the four shells at a correlation of 0.987
to 0.992, and 3.3\% of voxels exceed twice their certified floor. The replay sits 0.005 to 0.009 of $S_0$ above the
dataset on every shell, the area DiSCo's twelve-sided tubes lose against the round tubes simulated here
(Supplement~\ref{app:disco}). Replaying the 364-measurement protocol over the whole phantom takes 14~s on one GPU,
18~s with every tissue tier at 3~T, through the time-segment layout of Sec.~\ref{sec:results:pages}.

With the catalogue's white matter at 3~T, the pools' $T_2$, the walls' relaxivity and the sheath's $\chi$, the same
protocol replayed from the same pack lowers the direction-averaged shells by
0.004 to 0.007 of $S_0$ and single measurements by up to 0.034. The tiers act at one scale: in the voxel the
3~T tissue moves most at $b=1925$~s/mm$^2$, all of them together take the signal from 0.513 to 0.479, the $\chi$
alone by 0.020, and $T_2$ and the relaxivity by 0.007 each. At 7~T the catalogue gives both pools one $T_2$, and the
shells move by less than 0.001 while single measurements still move by 0.05.

Figure~\ref{fig:disco} and Table~\ref{tab:disco} are the tractography question, through one pipeline,
constrained spherical deconvolution \citep{Tournier2007} and probabilistic tractography, both dipy's
\citep{Garyfallidis2014}, whose ceiling the dataset's own images set at 0.905. The replay of the dataset's own scheme scores $0.911\pm0.002$ over
three noise seeds: the partitioned phantom is the ground truth the dataset was, to within the seed. The five
acquisitions the dataset never simulated separate by their diffusion contrast, not by their direction count: the
high-$b$ single shell scores as the dataset's own scheme with a sixth of its directions, above the three-shell
research scheme; the clinical shell at the dataset's diffusivity attenuates as $b\approx300$~s/mm$^2$ does in
vivo and gives the deconvolution little to work with; and the cosine OGSE at 50~Hz scores as the clinical shell at
the same $b$ but places its fibre peaks twice as far from the strands, $18^\circ$ against $10^\circ$, while
the stimulated echo at the same $b$, its diffusion time stretched to 45.9~ms by the mixing time, scores above both
at the clinical shell's peak error, questions the dataset could not have been asked. Against noise the research scheme scores 0.850 at SNR 10 and 0.894 at
SNR 50 (panel D), each point one more pass of the image loop over the same simulation.

The same simulation answers a $b$-tensor encoding (panel E). In the densest quarter of strand voxels the fractional
anisotropy falls from 0.47 where the strands run in one direction to 0.31 where they cross in three or more, while
the microscopic anisotropy \citep{Lasic2014mfa,Szczepankiewicz2015mfa} stays at 0.59 to 0.62: crossing lowers the
voxel's anisotropy, not the strands'.

\begin{table}[!htbp]
\centering\footnotesize
\caption{Six acquisitions replayed from the one DiSCo simulation at SNR 30 (three noise seeds) through one
tractography pipeline: the connectome score against the ground truth and the median angle between the
estimated FOD peak and the strands' mean orientation per voxel. The dataset's own images through it give the ceiling.}
\label{tab:disco}
\begin{tabular}{@{}lrrr@{}}
\toprule
Acquisition & Directions & Score & Peak error \\
\midrule
Dataset's images, noise-free / SNR 30 & 360 & 0.904 / 0.905 & $5.7^\circ$ / $7.7^\circ$ \\
Dataset's scheme, replayed & 360 & $0.911\pm0.002$ & $6.6^\circ$ \\
Connectome~2.0, $b=6000$~s/mm$^2$ & 60 & $0.911\pm0.001$ & $7.2^\circ$ \\
Research, $b=1000$, 2000, 3000 & 270 & $0.886\pm0.004$ & $6.3^\circ$ \\
Clinical, $b=1000$ (order 6) & 30 & $0.776\pm0.008$ & $9.8^\circ$ \\
Cosine OGSE 50~Hz, $b=1000$ & 60 & $0.769\pm0.009$ & $18.0^\circ$ \\
PGSTE, mixing time 38.3~ms, $b=1000$ & 60 & $0.803\pm0.003$ & $9.6^\circ$ \\
\bottomrule
\end{tabular}
\end{table}

\subsection{Anyone can query the phantom in a browser}\label{sec:results:pages}

\begin{figure}[!htbp]
\centering
\includegraphics[width=\linewidth,height=0.78\textheight,keepaspectratio]{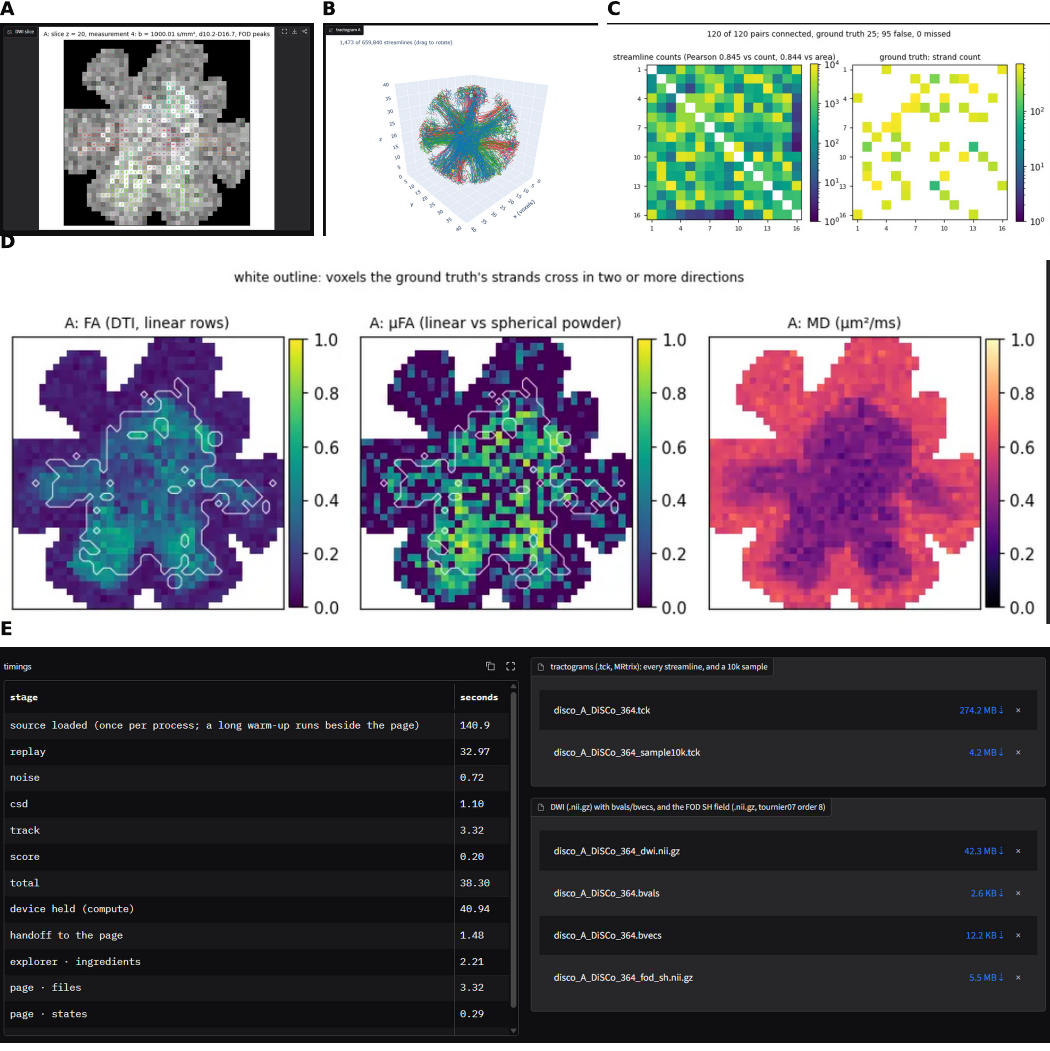}
\caption{The DiSCo page \citep{DiscoZeroSpace2026} as a visitor sees it, snapshots of the live page
(\texttt{figures/fig\_pages.py}). After a run of DiSCo's own 364-measurement protocol at SNR 30 with every tissue
tier at 3~T: (A) a slice of the replayed images with the fibre orientations, (B) the tractogram, (C) the connectome
against the strands' count, scored against the ground truth. (D) After a $b$-tensor run: fractional anisotropy,
microscopic anisotropy and mean diffusivity, the ground truth's crossing voxels outlined. (E) Each stage's time and
the files the visitor downloads: the images as NIfTI with their $b$-values and directions, the fibre orientations
and the tractogram.}
\label{fig:pages}
\end{figure}

DiSCo, the largest Monte Carlo phantom built for diffusion tractography, can now be imaged by anyone, with their own
acquisition, in a web browser \citep{DiscoZeroSpace2026}. On a public page a visitor designs a protocol, its shells
and directions on any of nine stored timings (PGSE, cosine OGSE, a stimulated echo and a $b$-tensor encoding),
chooses the tissue ($T_2$, surface relaxivity and the sheaths' susceptibility at 3 or 7~T) and a signal-to-noise
ratio, and presses one button. Within two minutes the page returns the whole phantom's Monte Carlo diffusion-weighted
images under that protocol, every tissue tier included, and for a diffusion protocol the fibre orientations, the
tractogram and the connectome scored against the ground truth, or for the $b$-tensor encoding the maps of mean
diffusivity and microscopic anisotropy (Fig.~\ref{fig:pages}); the visitor downloads the images as NIfTI with their
$b$-values and directions, and the tractogram. No simulation runs: every image is a replay of the one DiSCo walk, at
the accuracy its pack certifies, and the compute is the Hub's shared GPU pool's, charged to the visitor's own daily
quota (Table~\ref{tab:pagerun}).

\begin{table}[!htbp]
\centering\small
\caption{A visitor's run on the DiSCo page, on the Hub's shared GPU, every tissue tier at 3~T: GPU time per stage
and the time from the button to the results on the visitor's screen.}
\label{tab:pagerun}
\begin{tabular}{@{}lrrrrr@{}}
\toprule
Protocol & Replay & Maps or fit & Tracking & GPU in all & At the visitor \\
\midrule
DiSCo's own, 364 measurements, SNR 30 & 33~s & 1~s & 3~s & 41~s & 100~s \\
$b$-tensor, linear and spherical, SNR 100 & 25~s & 9~s & -- & 36~s & 72~s \\
\bottomrule
\end{tabular}
\end{table}

The page does not read the 294~GB pack but its time-segment layout: the pack contracted once per stored timing by
Eq.~\eqref{eq:shape} (Sec.~\ref{sec:methods:phantoms}), 169 bytes per walker and 51~GB in all
\citep{DiscoTimeSegmentsHub2026}, on which any $b$-value and direction is one phase per walker. The connectome
the page scores, 0.845, comes from \texttt{dmipy-fit}'s constrained spherical deconvolution and \texttt{dmipy-tract}'s probabilistic\footnote{\url{https://github.com/dmrai-lab/dmipy-tract}, on PyPI as \texttt{dmipy-tract}; AGPL-3.0.}
tracker, both on the GPU, in place of the dipy pipeline of Table~\ref{tab:disco}. What the page serves is what
Sec.~\ref{sec:results:phantoms} validated: the layout is within $2.2\times10^{-5}$ of the full replay, against a
combined bound of $3.0\times10^{-3}$, and the page's acceptance test refuses a build above $10^{-3}$. The price of
seconds is the timing: the page serves the timings it stores, and a new one is one more contraction of the pack,
done once rather than per visitor.

\section{Discussion}\label{sec:discussion}

Replay changes which part of the pipeline is scarce. Conventional Monte Carlo penalises a generator for its own
realism, ConFiG \citep{Callaghan2020config}, MEDUSA \citep{Ginsburger2019medusa}, CACTUS \citep{VillarrealHaro2023},
CATERPillar \citep{NguyenDuc2026} or an axon traced from microscopy \citep{Lee2020shape,Andersson2020axon}, because
every geometry sample must be simulated again for every question asked of it; replay simulates a sample once, to a declared accuracy, and every later question is an inner product.
The break-even is the second question: DiSCo is walked once, and every further acquisition is an image of the
whole phantom in seconds of a shared GPU (Sec.~\ref{sec:results:pages}). What this leaves untouched is the
parameter space, which Table~\ref{tab:knobs} says still costs one simulation per geometry. The factorisation has a precedent in
another field. Precomputed radiance transfer
\citep{Sloan2002prt} stores a spherical-harmonic record of light transport at every vertex of a scene once,
evaluates any lighting as an inner product with that record and rotates the record rather than the scene, as
replay rotates the acquisition rather than the simulation, and it turned global illumination from an offline computation into a real-time one;
replay does to spins what it did to light. The open problem moves upstream, to which tissue representations
are worth solving rather than how quickly a signal can be simulated: the generators above, and the measured
morphology they are built from \citep{Aboitiz1992,Liewald2014,Andersson2020axon}, become the scarce input, and
replay makes each of their samples worth more questions, not fewer. Three consequences follow for the field, and one
future direction.

\paragraph{Validation becomes a property of a file.} A simulator is trusted today by its paper: the
engine was checked against closed forms once \citep{HallAlexander2009camino,RafaelPatino2020mcdc,Kerkela2020disimpy},
and every result it later produces inherits that trust. A pack
carries its own certificate, each channel's error stated as terms with a closed form beside the Monte-Carlo floor, the
standard error of its walkers (Sec.~\ref{sec:theory:band}), so the claim ``this substrate is solved to $5\times10^{-3}$ for every waveform a
Prisma can play'' is a checkable property of an object anyone can download, not of a program anyone must believe. Other engines' walks
enter as packs read by the same reader, so engines are compared walker by walker rather than paper by paper
(Sec.~\ref{sec:results:meshes}); the specification was written for exactly that. Every experiment in this paper reads a certificate before it reads a
signal, and I would argue that a simulated reference should carry one.

\paragraph{The acquisition-and-model framework becomes explorable.} Le Bihan's many-to-one mapping from
tissue to signal \citep{LeBihan2026} has been explored by acquiring: a new contrast means a new scan of a
subject whose tissue is not known. A stored substrate is the other experiment, the one where the tissue is
fixed and the acquisition varies without simulating again, and the DiSCo page (Sec.~\ref{sec:results:pages}) puts that experiment in a
visitor's hands; DiSCo's acquisitions in Table~\ref{tab:disco} are a first map of it. Sequence design against a
stored tissue follows directly \citep{Alexander2008design,Drobnjak2010waveform}: the worst-waveform search of
Sec.~\ref{sec:results:band} is already a search over a scanner's deliverable set against a pack, and what it
maximised here, the certificate's error, can as well be a contrast.

\paragraph{Learned estimators get a training set that spans acquisitions.} Microstructure is increasingly
estimated by models trained on simulated signals: random forests and networks trained on Monte Carlo walks with
exchange \citep{NedjatiGilani2017,Hill2021permeability}, networks for a diffusion--relaxation model
\citep{deAlmeidaMartins2021}, likelihood-free inference that learns a posterior from simulations
\citep{Cranmer2020sbi,Jallais2022lfi}, and neural emulators that stand in for a simulator altogether
\citep{Kasim2022emulators}. Their accuracy is bounded by the distribution they were trained on
\citep{Gyori2022training}, and a Monte Carlo training set is tied to the protocol it was simulated with. A pack
removes that tie: one simulation yields labelled signals for any acquisition, field strength and tissue parameter
inside its certificate, each exact to it, so a training set spans protocols and tissue at the cost of replays rather
than walks, and an emulator, where one is still wanted, is checked against the replay it approximates.

\paragraph{A commons of solved substrates.} What makes a pack shareable is the contract and the compression: a
specification that says what a substrate, a pack, an acquisition and a phantom are \citep{FickRpkSpec2026}, and a
size small enough to host. The Hugging Face Hub is the infrastructure that fits it, because it serves pinned
revisions and byte ranges, so a pack is not a file on one computer but a resource the whole community reads by the
bytes its question needs: solved once and replayed indefinitely for any later question. Replaying needs nothing
more than \texttt{dmipy-replay}, a reader with NumPy as its only dependency that replays any pack on the Hub. Nor is
a pack a finished deposit: its endpoints are exact and a walk is Markov, so it can be continued by one more
certified window without walking the first again (Sec.~\ref{sec:methods:pack}).

\paragraph{Limitations.}\label{sec:discussion:limits} Nothing in this paper is compared with a measured sample, the ex vivo tissue with known
histology that validation of microstructure imaging rests on \citep{Dyrby2018validation,FieremansLee2018cookbook}. What it establishes is the
method: every channel, from the substrate to the voxel signal, is checked against a closed form, an independent
computation or another simulator. Turning it into a prediction of a scan needs a sample whose substrate is known
independently of that scan. What changes the position update still costs a simulation: the
geometry, the diffusivity and permeability, and the kinetics of a bound pool (Table~\ref{tab:knobs}), with two
exact exceptions. In time, a walk of duration $T$ at $D$, read as one of duration $T/a$ with $a\ge1$, is the walk at
$aD$, every channel rescaled with it, and across a membrane the walk at $(aD,\,a\kappa)$ and at no other pair
\citep{Powles1992}; only faster, since replay's in-step error grows as $1/a$ and the pack's certificate does not
cover it. Water's $D$ rises 2.5\,\% per kelvin \citep{Holz2000}, so on an impermeable substrate this is a
temperature knob. In space, scaling every length by $s$, every pool's $D$ by $s^2$ and permeability, surface
relaxivity and binding by $s$ leaves $\kappa L/D$, the crossing probability and $DT/L^2$ unchanged, so the walk at
size $L$ is the walk at size $sL$ at $(s^2D,\,s\kappa)$, read at gradient $s\mathbf G$; exact both ways, but for
$s>1$ the gradient $s\mathbf G$ can leave the waveforms the pack was certified for. The occupancy, contact and field
channels are defined by the walk that accumulates them, so only an engine that implements those definitions writes
them; another engine's walk enters as positions alone. Every certificate in this paper is a Monte-Carlo
floor, so a claim below a pack's floor is not one it can make; and one comparison stands outside the noise:
on the CACTUS strands this engine's surface relaxation sits 0.3 to 0.6\% of the signal above MCMRSimulator's
(Sec.~\ref{sec:results:meshes}), although it matches the closed form on tessellated cylinders; the cause is not yet
known.

\paragraph{From the voxel to the image.}\label{sec:discussion:image} Replay here ends at each voxel's signal on an
ideal scanner. A real scanner's own terms, its gradients' nonlinearity, concomitant fields and readout, and a
voxel's distribution of fibre poses enter at the voxel through the replay equation (Sec.~\ref{sec:theory:invariant})
without touching a pack; composing them is the step from an ideal scanner to a real one and from a voxel's signal to
an image, and it needs calibrations against measured maps that are future work.

Nothing in the walk's projection is specific to MRI: any Monte Carlo whose observables are linear functionals of the
path can keep the path's coefficients on a basis in place of the path, and answer every later question from them.

\section{Conclusion}\label{sec:conclusion}

A Monte Carlo simulation of tissue does not have to be run again to be asked again. Everything a scanner can do
to a spin enters the signal through a few records of each walker's history, the stored record has a band set by
the scanner and not by the tissue, a rotated substrate is a rotated acquisition, and what a real scanner adds
enters at the voxel. Stored once as a certified pack, 210 times smaller than its trajectories and
5\,000 times at a timing, a simulation answers any acquisition by an inner product, in pores to the closed forms, on realistic axons to two other engines, and on DiSCo to the published images,
and a phantom built this way is a public page that anyone can query in seconds. What remains scarce is no
longer the simulation but the choice of what to simulate: which substrates, at which accuracy, are worth solving
once for everyone. The pack specification and the commons are offered so that the choice can be a community's.

\section{Data and Code Availability}\label{sec:data}

Every signal in this paper is simulated; no human or animal data were used. Every replay pack of the paper is public
on the Hugging Face Hub under the \href{https://huggingface.co/SubstrateCommons}{\texttt{SubstrateCommons}}
organisation \citep{CanonicalPoresHub2026,PermeableSlabHub2026,DiscoReplayHub2026,DiscoTimeSegmentsHub2026,CoefficientWalksHub2026},
free for research and non-commercial use under CC BY-NC-SA 4.0, with a commercial licence on request; each dataset
states the terms of its packs beside the licence of the source geometry they were simulated in, which the packs
never relicense. The DiSCo substrate, its images and its ground truth are the DiSCo authors' dataset
(\href{https://doi.org/10.17632/fgf86jdfg6.1}{Mendeley Data}, CC BY 4.0) \citep{RafaelPatino2021disco}.

The software is public under the \href{https://github.com/dmrai-lab}{\texttt{dmrai-lab}} organisation on GitHub and
on PyPI: the engine \dmipysim{}, \texttt{dmipy-fit}, \texttt{dmipy-tract}, \texttt{dmipy-design} and the page's
application \texttt{disco-space} under AGPL-3.0 or a commercial licence, and the reader \texttt{dmipy-replay} and the
replay-pack specification \citep{FickRpkSpec2026} under Apache-2.0. The DiSCo page is public at
\url{https://huggingface.co/spaces/rfick/disco-replay} \citep{DiscoZeroSpace2026}. The scripts behind every figure,
which read the packs from the Hub at pinned revisions, are published with the paper's source on publication.

\section{Ethics Statement}\label{sec:ethics}

This study involved no human participants, human data, or animal subjects, and
required no ethical approval.

\section{Author Contributions}\label{sec:credit}

Rutger Fick: Conceptualization, Methodology, Software, Validation, Formal analysis, Investigation, Data curation,
Visualization, Writing -- original draft, Writing -- review \& editing, Project administration.

\section{Funding Information}\label{sec:funding}

This work received no external funding. The author self-funded it.

\section{Declaration of Competing Interests}\label{sec:coi}

The author offers commercial licences for the software and the replay packs described here (Sec.~\ref{sec:data}).

\section{Declaration of Generative AI and AI-Assisted Technologies}\label{sec:ai}

The engine, the analyses and the scripts behind every figure were developed with an AI coding agent
(Claude, Anthropic) under my direction, and the manuscript was drafted and revised with it. I set every question and
design decision, reviewed every change, and take full responsibility for the content. Every claim is checked against
a closed form, an independent route or another engine (Sec.~\ref{sec:results}), and every figure names the script
that produced it.

\section{Acknowledgements}\label{sec:ack}

This work did not begin as a plan for Monte Carlo replay. When coding agents became capable of serious
engineering, I wanted to see whether that capability could be pointed at infrastructure for the field, as dmipy once
was, and replay grew out of asking, at every step, what more a simulation could give and how it could be done more
efficiently. It is the product of a collaboration between a human acting as research director and AI coding agents
that grew markedly more capable over the six months of the project. I thank Marco Pizzolato for our conversations
over those six months, and the authors of DiSCo, CACTUS, MCMRSimulator.jl and MC/DC for making their data and
software open, without which none of the comparisons here would exist. The carbon footprint of developing this
computational efficiency framework is deliberately excluded, for reasons of personal embarrassment.

\section*{Supplementary Material}
S1, the experimental details behind Table~\ref{tab:experiments}; S2, the fibre field and its long-range sum. They follow the references, with their own bibliography.

\printbibliography

\clearpage
\begin{refsection}
\section*{Supplementary Material for: Replayable Monte Carlo MRI}
\setcounter{section}{0}\setcounter{figure}{0}\setcounter{table}{0}\setcounter{equation}{0}
\renewcommand{\thesection}{S\arabic{section}}
\renewcommand{\thesubsection}{S\arabic{section}.\arabic{subsection}}
\renewcommand{\thefigure}{S\arabic{figure}}
\renewcommand{\thetable}{S\arabic{table}}
\renewcommand{\theequation}{S\arabic{equation}}
\section{Experimental details}\label{app:experiments}

The recipes behind Table~\ref{tab:experiments}, in its order; every figure's data script is named in its
caption.

\subsection{The band certificate}\label{app:band}
Four substrates are simulated for $T=100$~ms at a save step of 0.1~ms, 20\,000 walkers each, seed 7: free
diffusion and cylinders of 2 and 10~$\mu$m diameter along $z$ at $D_0=2\times10^{-9}$~m$^2$/s, and the CACTUS
bundle \texttt{bundle\_00000\_capped} \citep{VillarrealHaro2023} with all three pools at its own
$1.7\times10^{-9}$. Each is stored at $K=999$, lossless on its 1001 saves. For each substrate, scanner class
(Table~\ref{tab:scanners}) and $K\in\{8,16,\ldots,512\}$ the certificate of Sec.~\ref{sec:methods:band} is
searched by projected Adam ascent on the simulation's grid from 32 oscillatory starts over 300 steps
(\texttt{dmipy\_design.certify}), fitted on the first 10\,000 walkers and measured on the other 10\,000 against
$|S_K|(1-e^{-D_0b_{>K}})$, with $b_{>K}$ from the waveform's sine coefficients above $K$ and the free bridge
variance. PGSE ($\delta/\Delta=10/40$~ms at $b=1000$ and 3000~s\,mm$^{-2}$ where deliverable) and cosine OGSE at
50--1600~Hz at each class's maximum amplitude are read through the same slices; the 50 and 25~ms prefixes repeat
the sweep on three classes. A class's certified bandwidth $f_c$ is the crossing of its free-diffusion
certificate with $\varepsilon=5\times10^{-3}$, interpolated in log--log between the sampled $K$.

\begin{table}[h]
\centering\footnotesize
\caption{The scanner classes of Fig.~\ref{fig:band} and Table~\ref{tab:classes}: the per-axis amplitude and slew
rate the bound takes its worst case over, and their sources. The coefficient walks keep 2\,000 modes over 100~ms,
a band of 10~kHz above every class.}
\label{tab:scanners}
\begin{tabular}{@{}lp{3.6cm}rrp{6.2cm}@{}}
\toprule
Class & System & $G_{\max}$ (mT/m) & $S_{\max}$ (T/m/s) & Source \\
\midrule
Prisma & Siemens Prisma, XR gradients & 80 & 200 & vendor specification, ``achievable simultaneously along each axis'' \citep{SiemensPrisma2024} \\
Connectom & Siemens Connectom, AS302 & 300 & 62.5 & \citet{McNab2013}: 200~T/m/s hardware, ``de-rated to 62.5~T/m/s during the diffusion encoding to prevent physiological stimulation'' \\
Connectome 2.0 & MGH/Siemens head-only coil & 500 & 600 & design target \citep{Huang2021}, reached per axis on the built scanner \citep{RamosLlorden2026} \\
MAGNUS & GE head-only insert, 1~MVA prototype & 200 & 500 & \citet{Foo2020}, ``200~mT/m and 500~T/m/s on each axis''; the 2024 product is rated 300~mT/m, 750~T/m/s \\
BGA-9S & Bruker BioSpec gradient, 90~mm bore & 760 & 6840 & measured on system at $I_{\max}$, $U_{\max}$ \citep{BrukerAlmanac2011} \\
Micro2.5 & Bruker microscopy insert, 40~mm bore & 1500 & 15\,000 & 1.5~T/m per direction; slew rate derived from the stated 5--95\% rise time below 100~$\mu$s at 60~A \citep{BrukerMicro25} \\
\bottomrule
\end{tabular}
\end{table}

\subsection{Canonical pores}\label{app:pores}
The 600 published packs \citep{CanonicalPoresHub2026} are each walked from their own seed for
$T=100$~ms and stored at $K=256$, the surface channel at 8 modes; each walker count is set by a pilot of 8\,000
walkers for any refocused Connectome~2.0 waveform and relaxivity up to 3~$\mu$m/s. The reference is the matrix method under the
Robin condition \citep{CallaghanStepisnik1995,CoddCallaghan1999,Grebenkov2007} (\texttt{figures/pore\_oracle.py}),
its split step resolving the pore's slowest relaxation $a^2/D$ by sixteen sub-steps (at most 64), the waveform's
samples held over the echo time and not beyond it. The rows are PGSE at $\delta/\Delta$ of 5/20, 10/40, 10/60 and
25/55~ms at $b$ of 1000 and 3000~s/mm$^2$, with and without relaxivity; cosine OGSE at 50 to 800~Hz and 0.3~T/m; and
the free-induction decay at 10 to 100~ms at relaxivities of 1 and 3~$\mu$m/s. Fig.~\ref{fig:pores}D reads nested
prefixes of each pack's shuffled ensemble.

\subsection{The permeable slab}\label{app:slab}
The two 2~$\mu$m compartments of a closed 4~$\mu$m slab have $\kappa$ of 10 and 100~$\mu$m/s between them, $T_2$
of 80 and 20~ms, and one is magnetised at $\chi=-0.1$~ppm at 3 and 7~T, its Lorentz-corrected field $\chi B_0/3$
with $B_0$ in its plane; every walker starts on one side. One simulation of 40\,000 walkers over 100~ms stores both
channels at the band of Sec.~\ref{sec:results:band}; every row is compared at its standard error. The reference is the
one-dimensional Bloch--Torrey equation with $D\,\partial_x m = \kappa[m]$ at the membrane, by finite differences with
a Strang split (400 nodes, the diffusion half-steps in the eigenbasis of the transmission Laplacian, the 180 as a
conjugation); doubling the nodes and halving the step changes it below $10^{-5}$.

\subsection{Realistic axons}\label{app:meshes}
The substrate is the CACTUS bundle \texttt{bundle\_00000} \citep{VillarrealHaro2023}: 344 of its 366 myelinated
strands, each an inner and an outer triangle mesh ($1.5\times10^6$ triangles; 22 strands whose surface is not
watertight are dropped), in a 30~$\mu$m box, reflecting at its faces in this engine's pack. Intra- and extra-axonal water are at
$D=1.7\times10^{-9}$~m$^2$/s with $T_2$ of 50 and 55~ms; myelin water at 40\% density, $T_2=10$~ms,
$\chi=-0.1$~ppm, no diffusion. One GPU walk of 111\,011 walkers over $T=100$~ms carries every channel of
Eq.~\eqref{eq:channels}, stored at $K=256$. The myelin field's basis is rasterised on $618^3$ nodes, two across
the thinnest sheath, membership by ray parity per strand (25.31\% of the cell is sheath on the raster,
$25.27\pm0.07\%$ by sampling the specification). The contrasts of Table~\ref{tab:experiments} are perpendicular PGSE
($\delta/\Delta=10/40$~ms) and OGSE at $b=1000$~s/mm$^2$ under the Connectome~2.0 limits, a spin echo at six echo
times, and the gradient echo at TE 40~ms with isotropic myelin susceptibility.

The two references walk the intra-axonal pool strand by strand from shared start positions, each strand periodic
along the 30~$\mu$m cell, exact at zero permeability: MCMRSimulator.jl~1.1 \citep{Cottaar2026mcmr} at its default step and MC/DC~1.50 \citep{RafaelPatino2020mcdc} at 2.3~$\mu$s, one process and seed
per strand. Their PGSE rows are a bipolar trapezoid pair, $\delta/\Delta=10/40$~ms, readout at 61~ms, nominal $b$ of
500 to 3000~s/mm$^2$ (527 to 3412 with the ramps), without RF. Each walk is also a pack of 88 bands from
positions at 36.8~$\mu$s, with the engine, version, commit, step and seeds in the header. The comparison of
Table~\ref{tab:enginesgrid} reads all three packs under PGSE ($\delta/\Delta=10/40$~ms, TE 60~ms) at $b$ of 1\,000 to
10\,000~s/mm$^2$ and at 0, 30, 60 and $90^\circ$ to the bundle axis, 11 azimuths per angle; on that grid the 88-band
packs' truncation stays below $8\times10^{-5}$, the reading's standard errors above $1.3\times10^{-4}$.
MCMRSimulator's collision loss $e^{-x\sqrt{\Delta t}}$ per hit is the Robin condition at $\rho=x\sqrt{D/\pi}$,
checked on a sphere against Brownstein--Tarr to $5\times10^{-5}$. On the extra-axonal pool, 500 spins under a
three-hour cap on eight threads, MCMRSimulator reached 74 of the 100~ms and MC/DC completed 60 spins. Scripts,
packs and logs: \texttt{figures/mesh\_xengine/packs/}.

\begin{table}[!htbp]
\centering\small
\caption{The intra-axonal pool of the CACTUS bundle's 176 strands walked by three engines: MCMRSimulator and MC/DC
strand by strand from the same 40\,678 start positions, stored as packs of 88 bands; dmipy-sim as the
intra-axonal pool of the whole-bundle pack, 37\,458 walkers. The walk's cost, and at $b=3000$~s/mm$^2$ the pack's
signal through one reader and its difference from dmipy-sim's as independent walks, with one standard error.}
\label{tab:engines}
\begin{tabular}{llllcc}
\toprule
Engine & Version & Step & Hardware, walk & $S$ at $b=3000$ & minus dmipy-sim \\
\midrule
dmipy-sim & 2.2.0 & 2.8~$\mu$s & one H100, 523~s (whole bundle) & $0.9428\pm0.0006$ & --- \\
MCMRSimulator.jl & 1.1.0 & default, 3--45~$\mu$s & 8 CPU threads, 6\,354~s & $0.9422\pm0.0005$ & $-6.0\pm7.7\times10^{-4}$ \\
MC/DC & 1.50 & 2.3~$\mu$s & 8 CPU threads, 4\,919~s & $0.9415\pm0.0006$ & $-13.0\pm7.8\times10^{-4}$ \\
\bottomrule
\end{tabular}
\end{table}

\paragraph{The surface channel.} On closed 12- and 64-gon cylinders of radius 1 and 2.5~$\mu$m at the bundle's
step, the boundary local time is within $5\times10^{-4}$ of $DtS/V$ and the free decay within 1.7 standard errors
of Brownstein--Tarr at TE 10 to 100~ms and $\rho_2$ of 1 and 3~$\mu$m/s. On the strands the replay sits above MCMRSimulator by
$2.9$ to $3.9\times10^{-3}$ of the signal at $\rho_2=1$~$\mu$m/s and $5.8$ to $6.2\times10^{-3}$ at 3~$\mu$m/s over
$b$ of 500 to 3000~s/mm$^2$, and by $6.1$ to $6.4\times10^{-3}$ against MCMRSimulator at a 2~$\mu$s step on 26
strands.

\subsection{DiSCo, partitioned}\label{app:disco}
Each of DiSCo's 12\,196 strands \citep{RafaelPatino2021disco} is two tubes swept along its centreline, the axolemma
at the listed diameter and a sheath at a $g$-ratio of 0.7; intra-axonal water inside the first and extra-axonal
water outside the second, both at $0.6\times10^{-9}$~m$^2$/s, with no exchange. The walk holds
$3.21\times10^8$ walkers, planned per voxel and pool for a floor of 1.0\% within a 300~GB dataset, on the 31\,815
of the 64\,000 voxels that hold water. The dataset's acquisition is
one echo time of 53.5~ms, ideal pulses, the three upper shells at the ActiveAx timings of
\citet{Alexander2010} (10.2/16.7, 7.6/45.9 and 17.7/35.8~ms) and the $b=1000$~s/mm$^2$ shell, whose timing the
dataset does not state, at the first shell's. The other acquisitions of Table~\ref{tab:disco} are at the same echo time, each the waveform of one
timing class of the published time-segment layout \citep{DiscoTimeSegmentsHub2026}, so the public page replays
exactly these: a clinical shell ($b=1000$~s/mm$^2$, 30 directions, $\delta/\Delta=12/24$~ms), a research
multi-shell ($b=1000$, 2000 and 3000~s/mm$^2$, 90 directions each, 17/30~ms) and a Connectome~2.0 shell
($b=6000$~s/mm$^2$, 60 directions, 8/20~ms), all with square lobes, a cosine OGSE at 50~Hz and
$b=1000$~s/mm$^2$, one 20~ms period each side of the refocusing pulse, and a stimulated echo (PGSTE; $\delta=7.6$~ms,
mixing time 38.3~ms, $\Delta=45.9$~ms) at $b=1000$~s/mm$^2$ with the OGSE shell's 60 directions; Rician noise is
added at replay. All, and the dataset's images, go through
constrained spherical
deconvolution \citep{Tournier2007} at order 8, or the largest even order the sparsest shell supports, the response
from the single-fibre voxels; probabilistic tracking seeded four times in every voxel of the sixteen regions, at most
$30^\circ$ of turn per half-voxel step (both dipy, \citealp{Garyfallidis2014}); and the Pearson correlation of the
$16\times16$ matrix of streamline counts with the ground truth over the 120 pairs. The strands' orientation in a
voxel is the principal axis of the length-weighted dyadic of their tangents.

DiSCo's tubes are meshes whose twelve-sided cross-section, inscribed in the nominal circle, holds
$\tfrac{12}{2\pi}\sin\tfrac{2\pi}{12} = 0.9549$ of its area, where the pack's tubes are round. Twelve straight tubes
(inner radii 1.0--2.4~$\mu$m, $g$-ratio 0.7, a 22~$\mu$m periodic cell, $4\times10^5$ walkers per pool, twenty
directions per shell) simulated round and twelve-sided through the dataset's acquisition differ by $+0.004$ to
$+0.008$ of $S_0$ on every shell, and the intra-axonal weight term alone reproduces that to within $10^{-3}$
(Table~\ref{tab:polygon}).

\begin{table}[t]
\centering\scriptsize\setlength{\tabcolsep}{4pt}
\caption{The same twelve tubes simulated round and twelve-sided through DiSCo's acquisition: direction-averaged
signal per pool and combined, the round-minus-polygon difference with its standard error, and its split into
the intra-axonal weight term and the two pools' own signal changes, all in units of $S_0$.}
\label{tab:polygon}
\begin{tabular}{r cc cc cc r r rrr}
\toprule
$b$ & \multicolumn{2}{c}{intra} & \multicolumn{2}{c}{extra} & \multicolumn{2}{c}{combined} & round $-$ poly & s.e.\ ($10^{-4}$) & weight & intra & extra \\
(s/mm$^2$) & round & poly & round & poly & round & poly & & & & & \\
\midrule
1\,000  & 0.8151 & 0.8163 & 0.6471 & 0.6466 & 0.7102 & 0.7064 & $+0.0038$ & 4.0 & $+0.0039$ & $-0.0004$ & $+0.0003$ \\
1\,925  & 0.6903 & 0.6924 & 0.4409 & 0.4399 & 0.5345 & 0.5288 & $+0.0057$ & 5.4 & $+0.0058$ & $-0.0007$ & $+0.0007$ \\
3\,094  & 0.6003 & 0.6015 & 0.2796 & 0.2776 & 0.3999 & 0.3916 & $+0.0083$ & 5.9 & $+0.0075$ & $-0.0004$ & $+0.0013$ \\
13\,191 & 0.2868 & 0.2891 & 0.0303 & 0.0277 & 0.1265 & 0.1197 & $+0.0068$ & 4.9 & $+0.0060$ & $-0.0008$ & $+0.0017$ \\
\bottomrule
\end{tabular}
\end{table}

\section{The fibre field and its long-range sum}\label{app:field}

\paragraph{One fibre.}
The field of an infinite hollow cylinder of inner radius $a$ and outer radius $b$, its sheath
carrying $\chi_I\,\mathbf I + \chi_A(\mathbf n\mathbf n^{\mathsf T} - \mathbf I/3)$ with
$\mathbf n$ the radial director, is known in closed form
\citep{WhartonBowtell2012}; the pack stores it in the thirteen-channel basis of the
$k$-space kernel \citep{MarquesBowtell2005}, $\delta B/B_0 = \chi_I[m/3 - \mathbf H\!\cdot\!\mathbf
M_I\!\cdot\!\mathbf H] + \chi_A\,\mathbf H\!\cdot\!\mathbf M_A\!\cdot\!\mathbf H$, with $m$ the
sheath indicator and $\mathbf H$ the unit field direction, so that $B_0$, its orientation and
both susceptibilities are contractions at replay. In the fibre's own frame ($\mathbf u$ the
axis, $\mathbf e_1$ the radial unit vector, $\mathbf e_2 = \mathbf u\times\mathbf e_1$,
$\mathbf P = \mathbf I - \mathbf u\mathbf u^{\mathsf T}$, $\mathbf S = \mathbf e_1\mathbf
e_1^{\mathsf T} - \mathbf e_2\mathbf e_2^{\mathsf T}$) the lumen has $\mathbf M_I = 0$ and
$\mathbf M_A = \tfrac12\ln(b/a)\,\mathbf P$, uniform, and the outside has $\mathbf M_I =
-\tfrac{b^2-a^2}{2r^2}\mathbf S$ and $\mathbf M_A = \tfrac{b^2-a^2}{12r^2}\mathbf S$, fitted
region by region to a two-dimensional $k$-space solve of one annulus and agreeing with it on every component.

\paragraph{A curved fibre.}
A strand is a polyline, and a segment's field is the dipole kernel integrated along it in closed form,
$\mathbf L = \int_{z_1}^{z_2}\!(\mathbf I - 3\,\mathbf n\mathbf n^{\mathsf T})/(4\pi r^3)\,\mathrm dz'$, with
$\mathbf n$ the unit vector from the source point to the field point and $z'$ the axial coordinate from the
point's foot on the axis: a segment of cross-section $A = \pi(b^2-a^2)$ carries $\mathbf M_I = A\,\mathbf L$ and
$\mathbf M_A = -A\,\mathrm{sym}(\mathbf L\mathbf T)$, with $\mathbf T = \mathbf P/2 - \mathbf I/3$ the
annulus-averaged director tensor; on a straight strand, however cut, this is the outside formula above. Within a strand's outer tube and one radius beyond
it, the strand's own term is its nearest segment's whole cylinder, and every other segment is seen from outside
its sheath, so that $\operatorname{tr}\mathbf M_I = m$ holds everywhere. Table~\ref{tab:field} is the cost of this rule at a joint; on one kinked strand of 20~$\mu$m
segments the segment sum agrees with the independent check below, on a 1~$\mu$m grid, to 0.1--0.2\% of the field's r.m.s.\ at 15 to
70~$\mu$m on every channel.

\begin{table}[!htbp]
\centering\small
\caption{The fibre field against the rasterised $k$-space route on one sheathed strand with
joints of $20^\circ$ (DiSCo's median) and $54^\circ$ (its sharpest), as the r.m.s.\ error over
each region in percent of the component's maximum; the sum of segment fields and, for comparison, the
nearest-segment rule.}
\label{tab:field}
\begin{tabular}{llcccc}
\toprule
joint & rule & lumen & sheath & within $3b$ & beyond \\
\midrule
$20^\circ$ & segment fields & 5.0 & 3.6 & 0.8 & 0.7 \\
           & nearest segment & 4.8 & 3.5 & 1.4 & 0.9 \\
$54^\circ$ & segment fields & 8.9 & 8.4 & 3.6 & 0.6 \\
           & nearest segment & 9.0 & 9.3 & 4.6 & 1.4 \\
\bottomrule
\end{tabular}
\end{table}

\paragraph{The long-range sum.}
Every segment's term is split by a cubic smoothstep $S(d)$ of its distance, from 0 at 11~$\mu$m to 1 at
18~$\mu$m, beyond the largest sheath's gate plus two cells. The near part $(1-S)$ is summed in closed form at the
walker's position; the far part $S$ is tabulated once from all 636\,951 segments on $401^3$ nodes of 2.5~$\mu$m
(1.7~GB in half precision, shipped with the substrate) and read by tricubic interpolation. On a 100~$\mu$m crop of
DiSCo's centre the read error is 0.98\% of the far part, 0.08\% of the whole field. The sum within 100~$\mu$m of a point differs from the sum within 50~$\mu$m by 8.1\% of
the isotropic channels' spread and 3.4\% of the anisotropic.

\paragraph{The independent check.}
The far part depends only on each segment's moment, so a source conserving it serves the $k$-space route:
cloud-in-cell assignment gives every cell of a 2.5~$\mu$m grid the sheath volume of the strand length it holds,
and that volume times $\mathbf T$, smoothed by a one-cell Gaussian, under the kernel on a box zero-padded to twice
the domain; it shares no closed form with the sum. On DiSCo's whole tract ($400^3$ cells) it agrees with the near
sum plus far grid at 29\,021 random points at least 15~$\mu$m from every segment to 0.15\% of the isotropic
channels' r.m.s.\ and 0.04\% of the anisotropic, correlation 1.0000, after the least-squares quadratic in
position, the block's own demagnetising field, is removed from both (with nothing removed, 1.5 and 0.1\%).

\printbibliography
\end{refsection}
\end{document}